\documentclass[]{iopart}

\usepackage{graphicx}
\usepackage{dcolumn}
\usepackage{bm}
\usepackage[sort&compress,numbers]{natbib}

\begin{document}

\title{Far-infrared spectra of lateral quantum dot molecules}

\author{M. Helle, A. Harju, and R. M. Nieminen} 

\address{Laboratory of Physics, Helsinki University of Technology,
P. O. Box 4100 FIN-02015 HUT, Finland }

\begin{abstract}

We study effects of electron-electron interactions and confinement
potential on the magneto-optical absorption spectrum in the
far-infrared range of lateral quantum dot molecules. We calculate
far-infrared (FIR) spectra for three different quantum dot molecule
confinement potentials.  We use accurate exact diagonalization
technique for two interacting electrons and calculate
dipole-transitions between two-body levels with perturbation
theory. We conclude that the two-electron FIR spectra directly
reflect the symmetry of the confinement potential and interactions
cause only small shifts in the spectra. These predictions could be
tested in experiments with nonparabolic quantum dots by changing the
number of confined electrons.
We also calculate FIR spectra for up to six noninteracting electrons and
observe some additional features in the spectrum.

\end{abstract}


\maketitle

\date{\today}

\section{Introduction}

Far-infrared (FIR) magneto-optical absorption spectroscopy is one
experimental technique to study electrons confined in semiconductor
quantum dots (QDs)~\cite{HeitmannPhysToday93,JacakHawryWojs}. It was,
however, realized at the early stage of QD research that FIR
spectroscopy is unable to reveal interesting many-electron effects in
parabolic-confinement QDs.  This is because the electromagnetic waves
couple only to the center-of-mass variables of electrons.  The
resulting FIR spectrum is rather simple, showing two branches,
$\omega_\pm$, as a function of magnetic
field~\cite{Sikorski,FIRMeurerPRL92}.  These branches, one with
positive energy dispersion ($\omega_+$) and one with negative energy
dispersion ($\omega_-$) as a function of magnetic field, are called
the Kohn modes.  The spectrum does not depend on the number of
electrons nor on the interactions between them. This condition in
parabolic QDs is called the generalized Kohn
theorem~\cite{MaksymChakrabortyPRL90}.  The condition can be lifted if
the confinement is not parabolic, and many experiments show more
complex FIR
spectra~\cite{HeitmannPhysToday93,FIRDemelPRL90,HochgrafePRB00,KrahnePRB01}.
Also calculations
~\cite{MeriPRL,MeriPhysica,Veikko_FIR_PRB91,Pfannkuche_FIR_PRB91,Madhav_FIR_PRB94,Magnusdottir_FIR_PRB99,Ullrich_FIR_PRB00,Gudmundson_FIRcondmat01}
have shown non-trivial FIR spectra of non-parabolic QDs. It is also
possible that spin-orbit interaction~\cite{ChakraPietil_FIR_PRL05} and
impurities near quantum dots~\cite{Jaime_FIR_PRB02} can have an effect
on the FIR spectrum. In a nonparabolic QD, the relative internal
motion of electrons could be accessible with the FIR spectroscopy, but
recent studies suggest that additional features in the FIR spectra are
still of collective nature~\cite{MeriPRL}. The interpretation of the
observed FIR spectra of a nonparabolic QD is usually far from trivial.
It is clear that deviations arise from nonparabolic confinement, but
the detailed cause of the deviations, thus the interpretation of
spectrum, is not always straightforward. It is especially interesting
to see how many-electron interactions appear in the FIR spectra of
QDs.

In the calculation of FIR spectra of two-electron lateral double
QD~\cite{MeriPRL} and lateral four-minima quantum dot molecule
(QDM)~\cite{MeriPhysica} it was shown that the deviations from the
Kohn modes arise mainly from nonparabolic confinement and interactions
have only minor effects on the spectra. Effects of relative motion of
electrons on FIR spectra were studied by turning electron-electron
interactions on and off.  The results indicate that the FIR spectra of
QDMs reveal mainly the center-of-mass collective excitations of
electrons. Actually more pronounced deviations from Kohn modes are
observed when interactions between the electrons are turned off. This
is an interesting result, suggesting that less features are observed
with more electrons in a non-parabolic QD. These studies, however,
include only two electrons and the FIR spectra may depend on particle
number. Also as more single-particle levels are occupied the
center-of-mass excitations with slightly different energies may be
observed where interactions do not, necessarily, have any significant
role. Calculations where electron interactions can be turned on and
off could show how the electron-electron interactions show up in FIR
spectra for greater electron numbers ($N > 2$).

In this article we study three different lateral quantum dot molecule
(QDM) confinements. We study two-minima QDM (double dot), and
square-symmetric and rectangular-symmetric four-minima QDMs. We
calculate the FIR spectra for all the QDM confinements and compare them to the
parabolic-confinement QD FIR spectrum. We analyze in detail the effect
of electron-electron interactions on the FIR spectra. We use very
accurate exact diagonalization technique for interacting electrons but
limit our studies to two interacting electrons. We also calculate
the FIR spectra of non-interacting electrons up to six electrons by
occupying single-particle levels and calculating dipole-transition
elements between the single-particle levels. These provide some
insight how FIR spectra are modified with greater electron numbers,
even if we do not include electron-electron
interactions. Especially if excitations are collective also with the
greater electron numbers, the FIR spectra are basically given by the
transitions between single-particle levels of the quantum dot
confinement in question.

\section{Model and method}  
\label{Model}

\begin{figure}
\hfill
\includegraphics*[width=0.5\columnwidth]{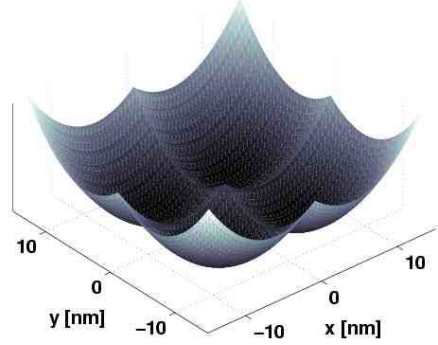}
\caption{Confinement potential of square-symmetric ($L_x=L_y=5$ nm)
four-minima quantum dot molecule.}
\label{pot}
\end{figure}

We model the interacting two-electron quantum-dot molecule with a
two-dimensional Hamiltonian
\begin{equation}
H = \sum _{i=1}^2\left ( \frac{ ( {- i {\hbar} \nabla_i}
-\frac ec \mathbf{A})^2 }{2 m^{*}} + V_\mathrm{c}({\bf
r}_{i}) \right ) +  \frac {e^{2}}{ \epsilon   r_{12} } \ ,
\label{ham}
\end{equation}
where $V_\mathrm{c}$ is the external confinement potential of the QDM
taken to be
\begin{equation}
 V_\mathrm{c}({\bf r}) = \frac 12 m^* \omega_0^2 \min \left\{ \sum_j^M ({\bf
 r} - {\bf L}_j)^2 \right\}.
\end{equation}
The coordinates are in two dimensions ${\bf r} = (x,y)$, the ${\bf
L}_j$'s (${\bf L}_j = (L_x,L_y)$) give the positions of the minima of
QDM potential, and $M$ is the number of minima. When ${\bf L}_1=(0,0)$
(and $M=1$) we have a single parabolic QD. With $M=2$ and ${\bf
L}_{1,2} = (\pm L_x,0)$ we get a double-dot potential. We also study
four-minima QDM ($M=4$) with minima at four possibilities of $(\pm
L_x,\pm L_y)$ (see Fig.~\ref{pot}). We study both square-symmetric
($L_x=L_y$) and rectangular-symmetric ($L_x \neq L_y$) four-minima
QDMs.  The confinement
potential can also be written using the absolute values of $x$ and $y$
coordinates as
\begin{eqnarray}
V_\mathrm{c}(x,y) &=& \frac 12 m^* \omega_0^2 \times \nonumber\\ &&
\left[ r^2 - 2 L_x |x| - 2 L_y |y| + L_x^2 + L_y^2 \right] \ .
\label{Vc_auki}
\end{eqnarray}
For non-zero $L_x$ and $L_y$, 
the perturbation to the parabolic potential comes from the linear
terms of $L_x$ or $L_y$ containing also the absolute value of the
associated coordinate.

We use the GaAs material parameters $m^*/m_e=0.067$ and
$\epsilon=12.4$, and confinement strength $\hbar\omega_0=3.0$
meV. This confinement corresponds to harmonic oscillator strength of
$l_0 = \sqrt{\hbar/\omega_0 m^*} \approx 20$ nm. We concentrate on
closely coupled QDMs where $L_{x,y} \leq l_0$. The magnetic field (in
$z$ direction) is included in the symmetric gauge by vector potential
$\mathbf{A}$. The Hamiltonian of Eq. (\ref{ham}) is spin-free, and the
Zeeman energy can included in the total energy afterwards ($E_Z =
g^*\mu_B B S_Z$ with $g^* = -0.44$ for GaAs). We disregard the
threefold splitting of each triplet state ($S_Z=0,\pm 1$) and consider
only the lowest energy one ($S_Z=1$).

We exclude the explicit spin-part of the wave function and expand the
many-body wave function in symmetric functions for the spin-singlet
state ($S=0$) and anti-symmetric functions for the spin-triplet state
($S=1$).
\begin{eqnarray}
 \Psi_S({\bf r}_1,{\bf r}_2) = \sum_{i \leq j} \alpha_{i,j} \{ 
 \phi_i({\bf r}_1)\phi_j({\bf r}_2) \nonumber \\
 + (-1)^S \phi_i({\bf r}_2)\phi_j({\bf r}_1) \},
\end{eqnarray}  
where $\alpha_{i,j}$'s are complex coefficients. 
The one-body basis functions $\phi_{i}({\bf r})$ are 2D Gaussians. 
\begin{equation}
\phi_{n_x,n_y}({\bf r}) = x^{n_x}y^{n_y} e^{-r^2/2}, 
\end{equation}   
where $n_x$ and $n_y$ are positive integers.  The complex coefficient
vector $\alpha_{i,j}$ of $\Psi_l$ and the corresponding energy $E_l$
are found from the generalized eigenvalue problem where the overlap
($S$) and Hamiltonian ($H$) matrix elements are calculated
analytically ($H {\bf \alpha}_l=E_l S {\bf \alpha}_l$). The matrix is
diagonalized numerically.

\begin{figure}
\begin{center}
\hfill
\includegraphics*[width=0.7\columnwidth]{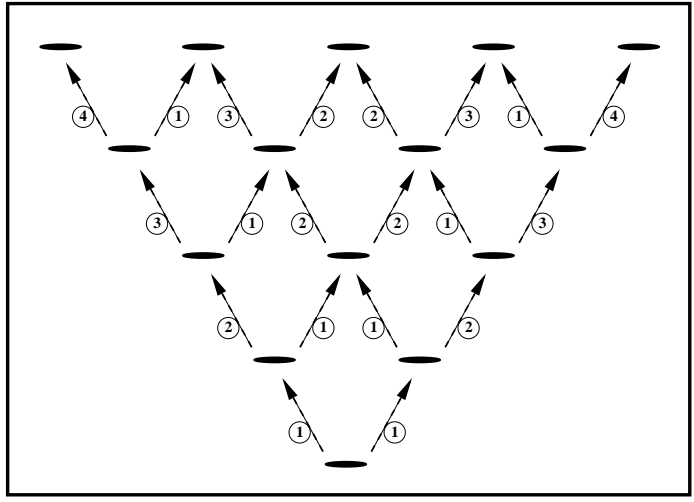}
\end{center}
\caption{Relative transition probabilities between Fock-Darwin energy
levels $P=\langle \phi_{n', l \pm 1} | e^{\pm i \phi} {\bf r} |
\phi_{n, l} \rangle$.}
\label{kokkokuva}
\end{figure}

The FIR spectra are calculated as transition probabilities from ground
state ($E_0$) to excited states ($E_l$) using the Fermi golden rule
within the electric-dipole approximation:
\begin{equation}
\mathcal{A}_{l \pm} \propto \left| \left< \Psi_{l} \left| e^{\pm i \phi}
\sum_{i=1}^{2} \mathbf{r}_i \right| \Psi_{0} \right> \right| ^2
\delta(E_l-E_0-\hbar \omega).
\end{equation} 
We assume circular polarization of the electromagnetic field: $e^{\pm
i \phi}\sum_i \mathbf{r}_i = \sum_i (x_i \pm i y_i) = z_{\pm}$, where
plus indicates right-handed polarization and minus left-handed
polarization. The results are presented for non-polarized light as an
average of the two circular polarizations.

\begin{figure}
\hfill
\begin{minipage}{0.25\columnwidth}
\includegraphics*[width=0.9\columnwidth]{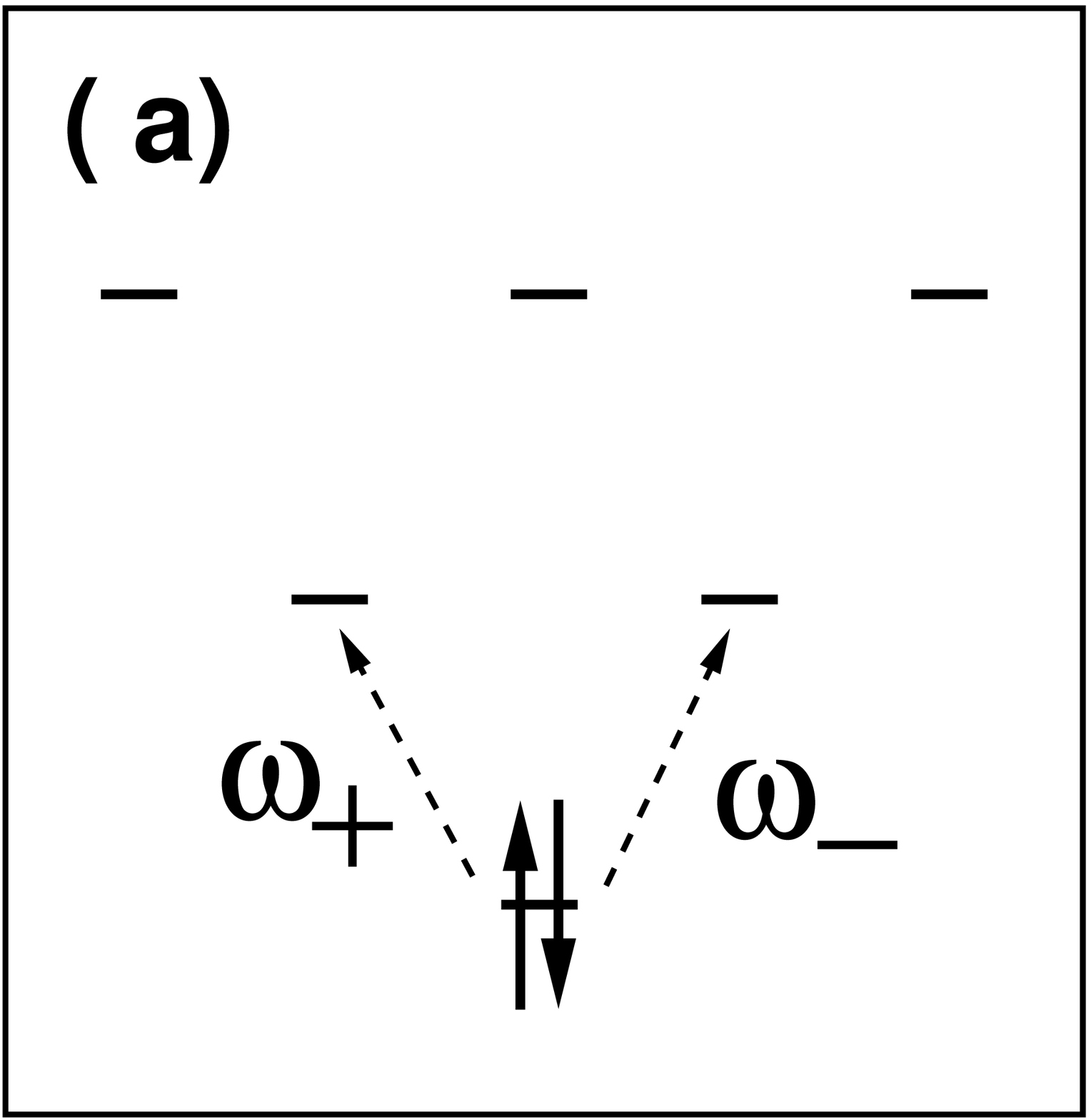}

\includegraphics*[width=0.9\columnwidth]{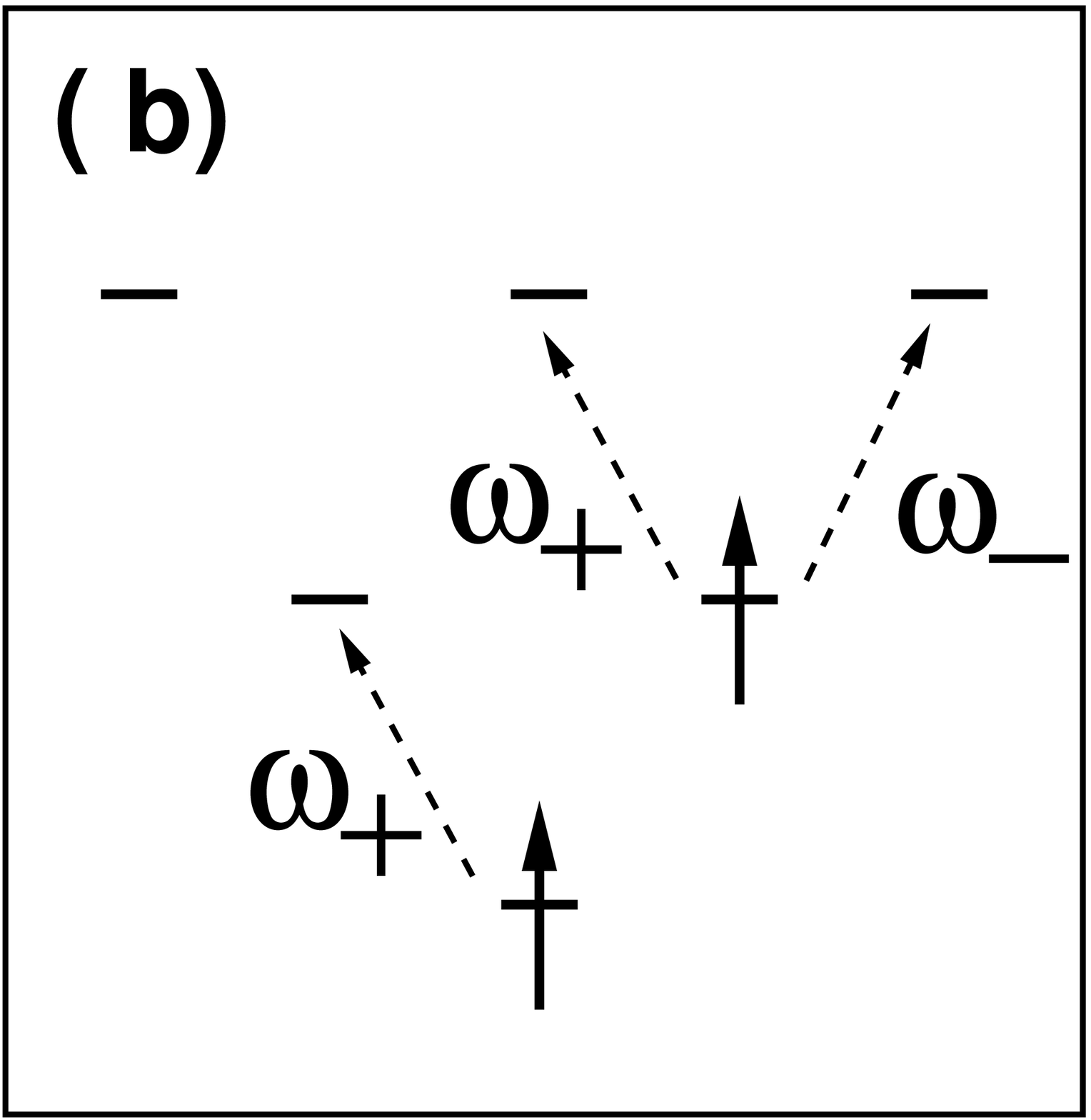}
\end{minipage}
\begin{minipage}{0.65\columnwidth}
\begin{center}
\includegraphics*[width=0.99\columnwidth]{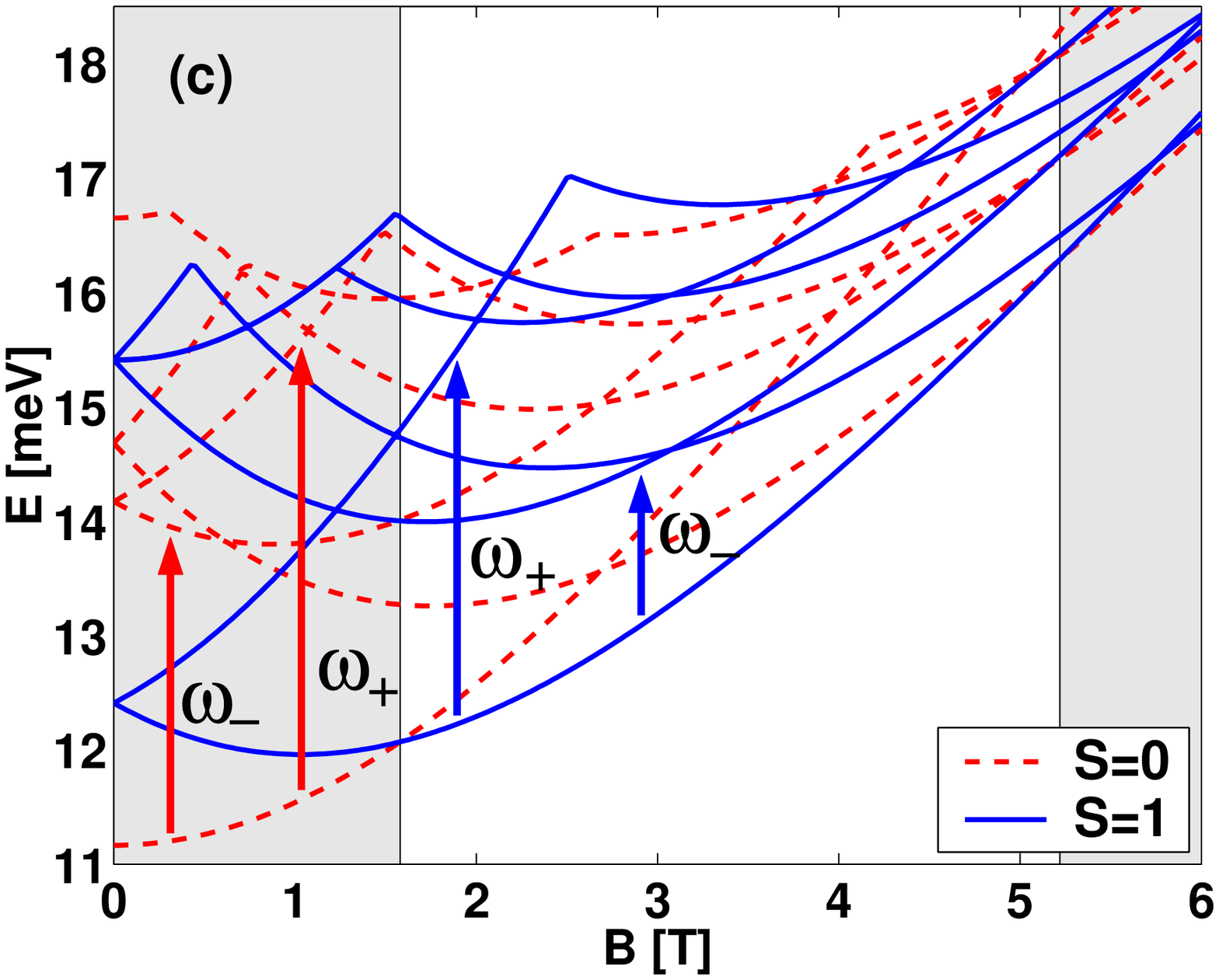}
\end{center}
\end{minipage}
\caption{Transitions for spin singlet (a) and spin triplet (b). (c)
  Two-particle energy levels of a parabolic quantum dot. Red lines
  show singlet energy levels and blue lines triplet energy
  levels. Gray background shows singlet ground state regions and white
  triplet ground state regions. Dipole transitions from one spin type
  to another are forbidden.}
\label{Kohnmodes}
\end{figure}

\section{Far-infrared spectra of two-electron quantum-dot molecules}
\label{FIR}

The dipole-allowed magneto-optical excitation spectrum of an isolated
harmonic-confined QD consist of two branches $\omega_+$ and
$\omega_-$, whose energy dispersion is well understood and does not
depend on the number of electrons in the quantum dot:
\begin{equation}
\Delta E_{\pm} = \hbar \omega_{\pm}=\hbar \sqrt{\omega_0^2 + (\omega_c/2)^2}
\pm \hbar \omega_c/2.
\end{equation}
$\omega_0$ describes the external confinement, $\omega_c=eB/m^*$ is
the cyclotron frequency, and $m^*$ is the effective mass of
electron. Fock-Darwin energy levels as a function of magnetic field
are given by $E_{nl} = (2n+|l|+1) \hbar \omega - \frac 1 2 l \hbar \omega_c$,
where $n=0,1,2,...$ and $l=0, \pm 1,...$ are principal and azimuthal
quantum numbers, respectively, and
$\omega=\sqrt{\omega_0^2+(\omega_c/2)^2}$. In the dipole-allowed
transitions between Fock-Darwin states angular momentum must change by
unity, $\Delta l= \pm 1$. Fig.~\ref{kokkokuva} shows relative
transition probabilities between Fock-Darwin energy levels $P=\langle
\phi_{n', l \pm 1} | e^{\pm i \phi} {\bf r} | \phi_{n, l} \rangle$.

For a two-electron QD the ground state can be either spin singlet
($S=0$) or spin triplet ($S=1$) depending on the magnetic field
strength. Even if many-body energy levels have more complicated
magnetic field dispersion than the single-particle levels, the
dipole-transitions in the many-body case always equal to transitions
between the single-particle levels in a \emph{parabolic}
confinement. A schematic picture of transitions at zero magnetic field
is shown in Fig.~\ref{Kohnmodes} (a) and (b) for spin-singlet and
spin-triplet states, respectively. Fig.~\ref{Kohnmodes} (c) shows a
few lowest two-body energy levels as a function of magnetic field for
the singlet and triplet states of a parabolic QD. The gray background
marks the magnetic field region of the spin-singlet ground state and
white the spin-triplet ground state. Dipole transitions from one spin
type to another are forbidden.

\subsection{Interacting two-electron spectra}

\begin{figure}
\hfill
\includegraphics*[width=0.4\columnwidth]{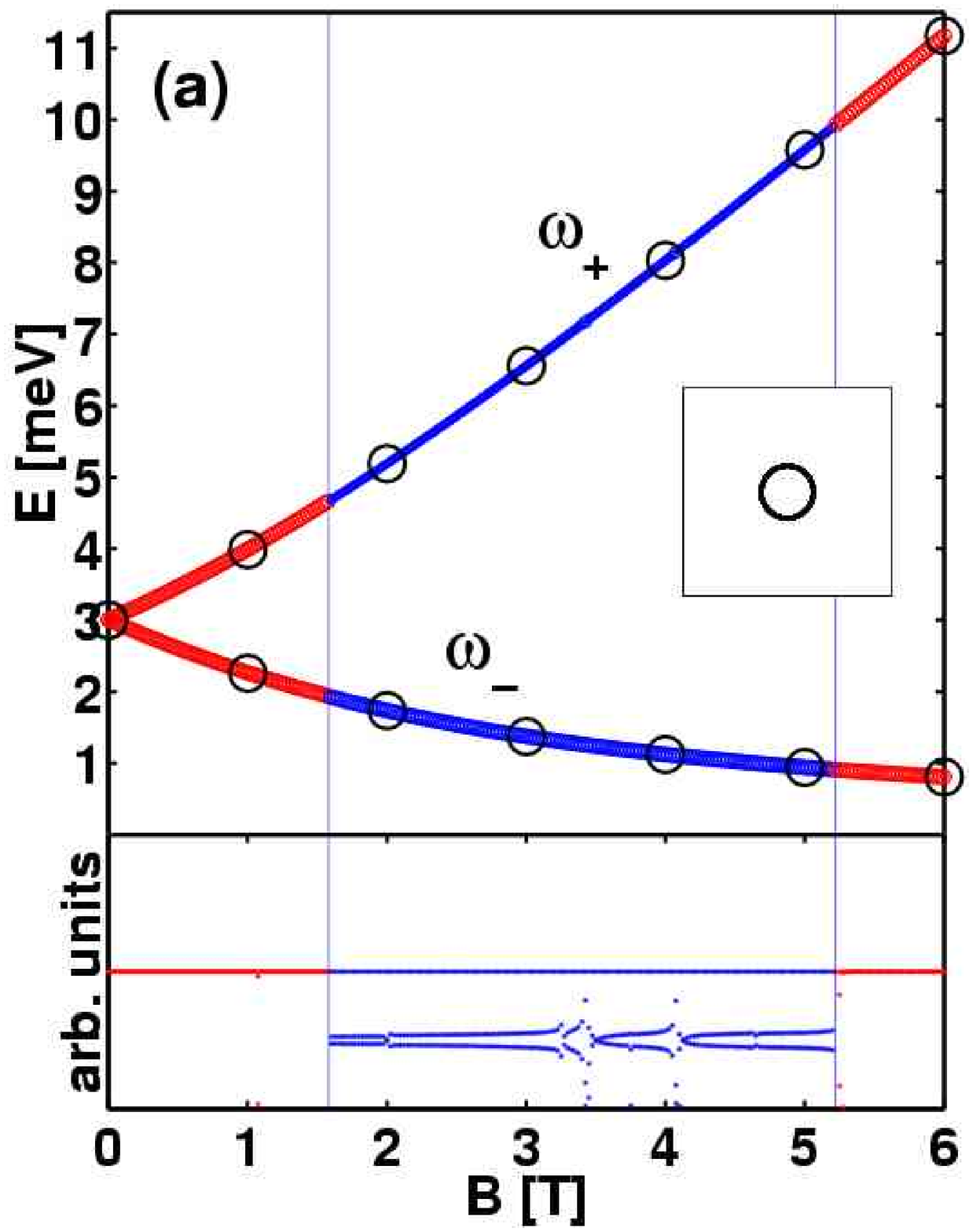}
\includegraphics*[width=0.4\columnwidth]{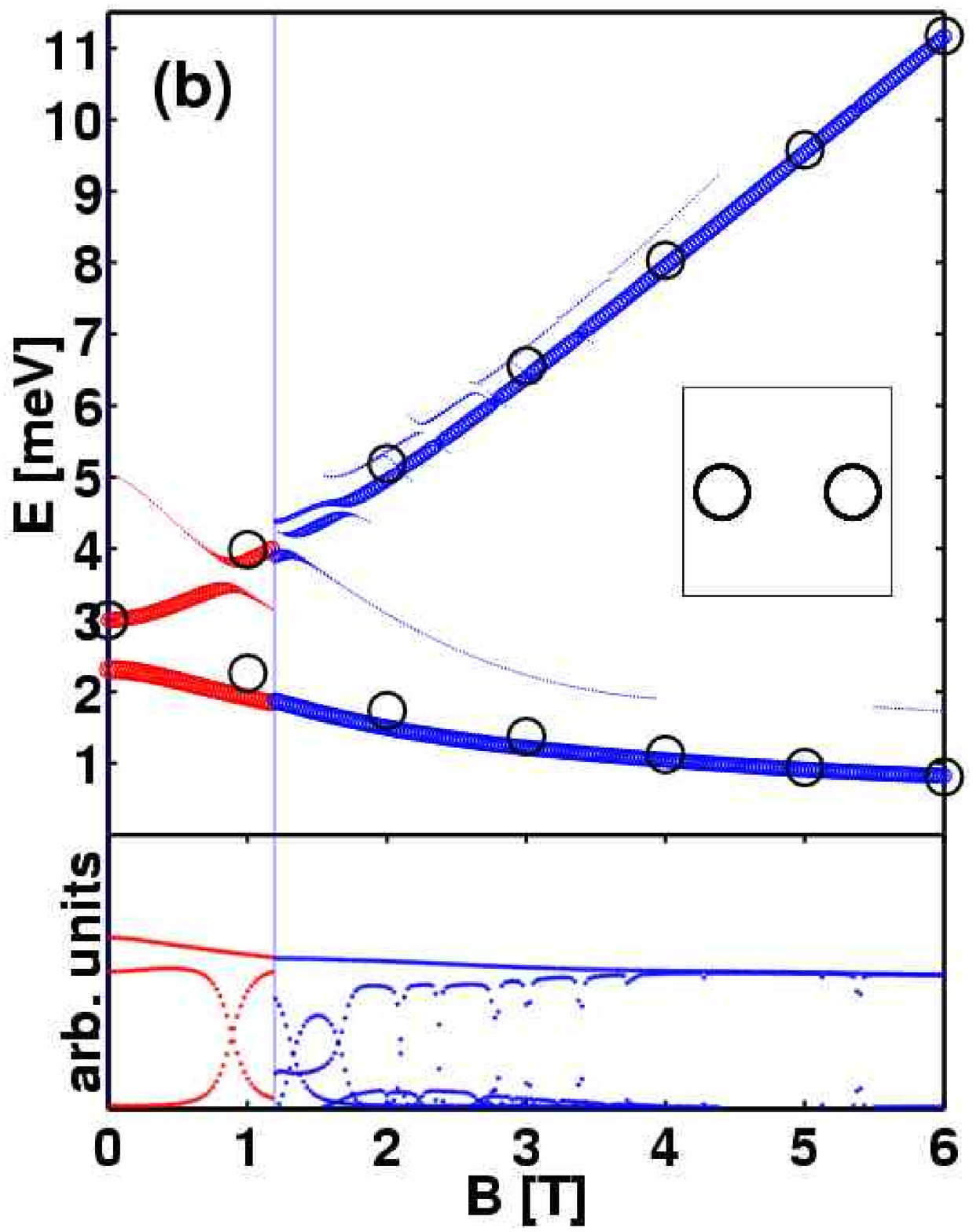}

\hfill
\includegraphics*[width=0.4\columnwidth]{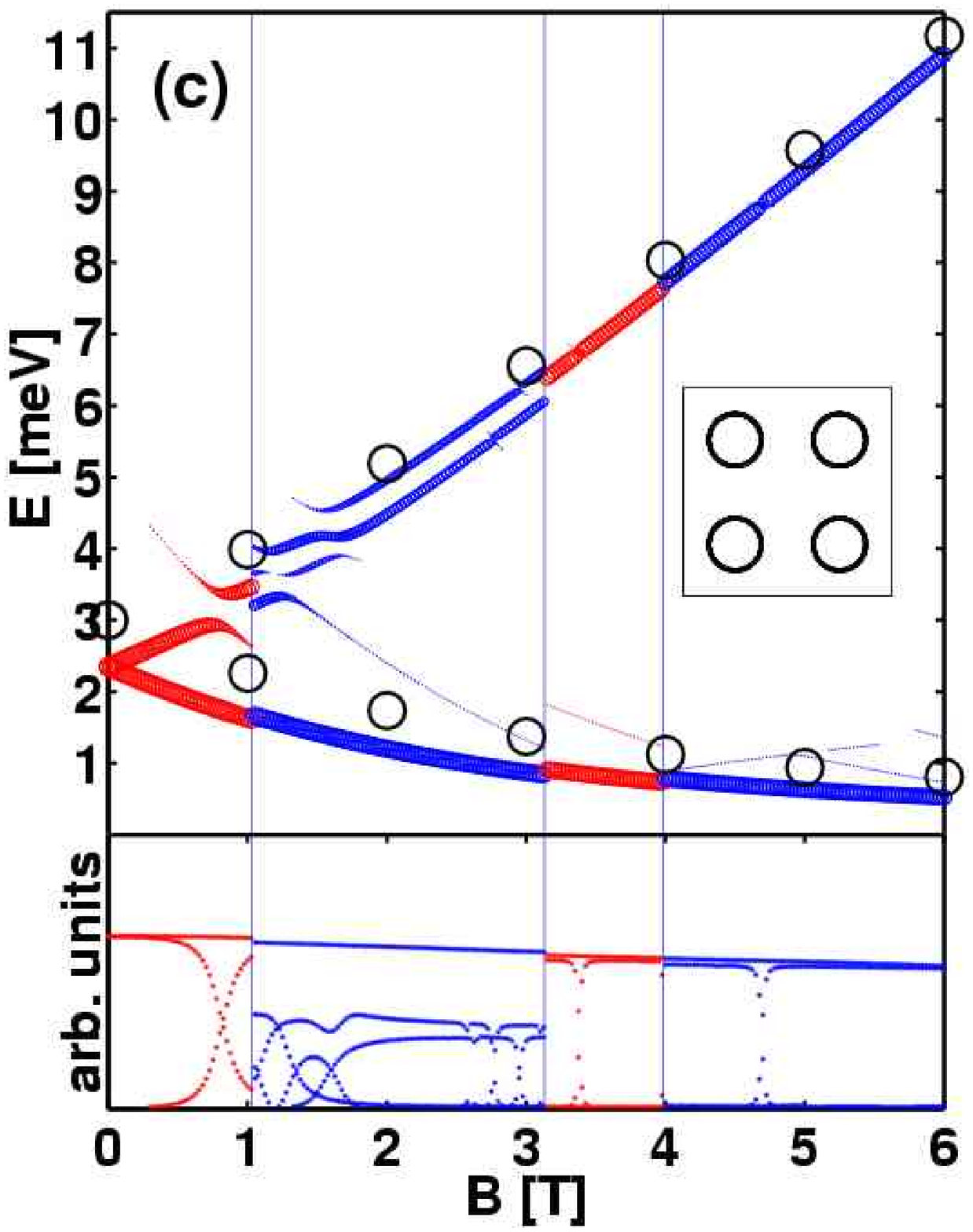}
\includegraphics*[width=0.4\columnwidth]{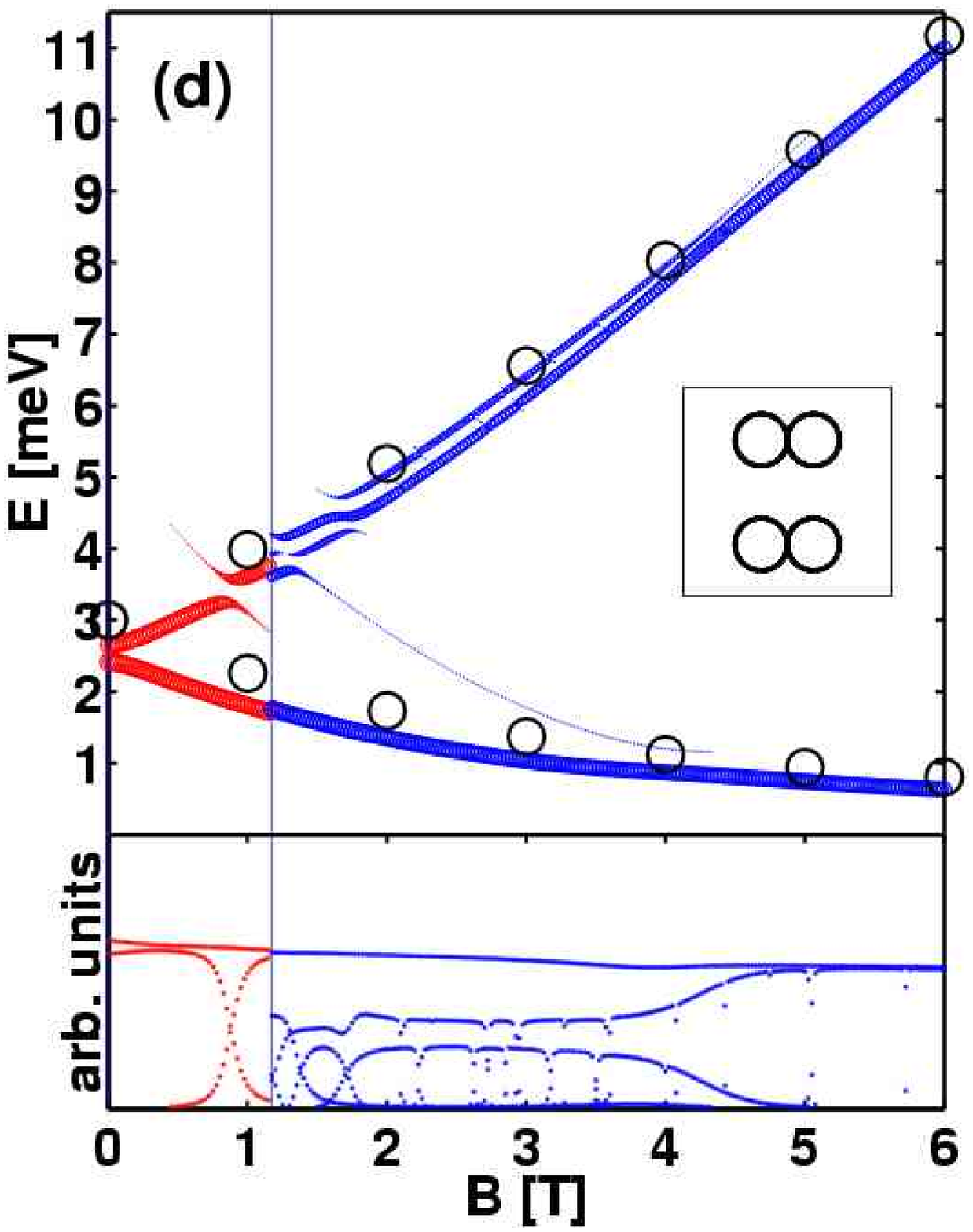}
\caption{Calculated far-infrared spectra of single parabolic quantum dot in
(a), double dot with $L_x=15$ nm in (b), square-symmetric four-minima
quantum-dot molecule with $L_x=L_y=10$ nm in (c), and rectangular-symmetric four-minima
quantum-dot molecule with $L_x=5,L_y=10$ nm in (d).  The energy of absorbed light is
given in meV as a function of magnetic field. The width of the line is
proportional to the transition probability, which is also plotted
below each spectrum (in arbitrary units) for each of the branches in
the spectrum. Transitions with small probability are excluded from the
spectra. Only the lines with transitions probability exceeding one
percent of the maximum transition probability are included in the
spectra. Large open circles in the spectra represent two Kohn modes
plotted with one tesla spacings. Vertical lines indicate the
singlet-triplet (or triplet-singlet) transition points.  }
\label{FIRcomb}
\end{figure}

Fig.~\ref{FIRcomb} shows the calculated far-infrared (FIR) absorption
spectra for two interacting electrons. Fig.~\ref{FIRcomb} (a) shows
the FIR spectrum for parabolic QD, (b) for two-minima QDM (double dot),
(c) for square symmetric four-minima QDM and (d) for
rectangular-symmetric four-minima QDM.  The energy of absorbed light
is given in meV as a function of magnetic field. The width of the line
is proportional to the transition probability, which is also plotted
below each spectrum (in arbitrary units) for each of the branches in
the spectrum. There are, of course, many more possible transitions
with nearly zero or small probability. Only the lines with transitions
probability exceeding $1$\% of the maximum transition
probability are included in Fig.~\ref{FIRcomb}. Large open circles in
the spectra represent the two Kohn modes, $\omega_+$ and $\omega_-$,
plotted with $1$T spacings. In QDMs there are ground state
transitions from spin-singlet ($S=0$) to spin-triplet ($S=1$) ground
states, and the other way round, as a function of magnetic
field~\cite{AriPRL02,MeriQDM1_PRB05}. Vertical lines indicate the
singlet-triplet (or triplet-singlet) transition points and the singlet
ground state regions of spectrum are marked with red color and the
blue lines denote triplet regions. Dipole transitions from one
spin type to another are forbidden.

Fig.~\ref{FIRcomb} (a) shows the calculated FIR spectra for a single
parabolic QD. The dipole transition probabilities are calculated from
the two-body ground state level to higher two-body energy levels as
shown in Fig.~\ref{Kohnmodes} (c). The spectrum of Fig.~\ref{FIRcomb}
(a) shows two Kohn modes $\omega_+$ and $\omega_-$. These coincide
perfectly with the open circles presenting Kohn modes, as they should,
since in a parabolic QD the FIR spectra does not depend on the number
of electrons in the QD nor on the interactions between
them~\cite{MaksymChakrabortyPRL90}. The magnitude of the absorbed
light can only change when the number of electrons in the QD is
changed~\cite{FIRMeurerPRL92,HeitmannPhysToday93}. However, in the
two-electron QD of Fig.~\ref{FIRcomb} (a) the width of the upper
branch is halved after the first singlet triplet transition. The
transition line is split into two degenerate transitions with almost
equal transition probabilities, which, if summed up, would equal to
the transition probability of $\omega_-$.  In the calculations we can
identify two degenerate levels to which electrons can be excited (see
Fig.~\ref{Kohnmodes} (b)), but in experiments the observed quantity is
the overall absorption of the energy in question. Therefore one can
observe Kohn modes with constant transition probabilities as a
function of magnetic field.

Fig.~\ref{FIRcomb} (b), (c), and (d) show FIR spectra for quantum-dot
molecules. Now as the symmetry of the confinement is lower, the center
of mass and relative motion do not decouple. There are clear
deviations from the Kohn modes. In all of the QDM potentials we can
see some general features. The main branches of the spectra always lie
below the Kohn modes (below open circles). There are clear
anticrossings and the upper branch is split to two in some parts of
the triplet spectra.

In the double dot spectra of Fig.~\ref{FIRcomb} (b), the excitations
at zero magnetic field are identified to a lower energy one along the
long axis (the line connecting two dots, $x$ axis) and a higher energy
one along the short axis ($y$ axis). This can be also verified by
calculating the absorption of linearly polarized
light~\cite{MeriPRL}. At $B=0$ the excitation of $\omega_+$ is
unaffected by the interactions between electrons and coincides with
the confinement ($\hbar \omega_0 = 3$ meV), because in the double dot
the potential is still parabolic along the $y$ axis. However, with
non-zero $B$ this is no longer true since a magnetic field mixes the
two linear polarizations. At high magnetic field the spectrum
approaches Kohn modes and no anticrossings are observed. In the high
field region the electrons are effectively more localized into
individual dots~\cite{MeriQDM1_PRB05}. Then the non-parabolic nature
of the confinement potential has a less important role, since the
electron density is more localized close to the parabolic minima.

After the singlet-triplet transition there appears an additional
level, $\omega_{+2}$, above $\omega_+$ with a lower transition
probability.  A simple explanation for two positive dispersion levels
would be in the picture of Fig.~\ref{Kohnmodes} (b) where the two
transitions of $\omega_+$ mode could have slightly different energy
difference in a non-parabolic confinement. However, the interpretation
is not so straightforward, since the dipole transitions are calculated
between two-body levels and not occupied single-particle levels, as we
will see when we analyze interactions in the next subsection.  Another
deviation from the Kohn modes of a parabolic QD are seen as small
discontinuities in the energy of the spectral lines when the singlet
changes to triplet. However, this discontinuity is very small and may
not be resolved in experiments. See Ref.~\cite{MeriPRL} for more
details on double dot FIR spectra.

In the square-symmetric four-minima QDM (Fig.~\ref{FIRcomb} (c)) the
potential is identical in $x$ and $y$ directions resulting in only one
excitation at $B=0$ as the long and short axes are equal. This
excitation is clearly lower than the $\hbar \omega_0 = 3$ meV
confinement energy of a single minimum. The upper branch is split to
two levels in the first triplet region at magnetic field values
between $1$ and $3$ T, but in the second triplet region $B > 4$ T
there is only one upper branch. In square-symmetric four-minima QDM the
two levels of the split-off upper branch have almost equal transition
probabilities.  Discontinuities in the spectra appear in the
singlet-triplet (or triplet-singlet) transition points. A brief
representation of the FIR spectra of square-symmetric four-minima QDM
can also be found in Ref.~\cite{MeriPhysica}.

The spectrum of rectangular-symmetric four-minima QDM, in
Fig.~\ref{FIRcomb} (d), has similar features as the double dot
spectra. There is a gap between the two branches at $B=0$.  The lower
branch at $B=0$ in Fig.~\ref{FIRcomb} (d) corresponds to excitation
along $y$ axis and the upper branch along $x$ axis. The upper mode
does not start at $3$ meV, as in the double dot, since the potential
is not parabolic in $x$ direction. After the singlet-triplet
transition an additional mode, $\omega_{+2}$, appears above
$\omega_+$. The transition probability of $\omega_{+2}$ is higher
compared to the $\omega_{+2}$ of double dot, but the transition
probability is not as high as in the square-symmetric four-minima
QDM. The modes of Fig.~\ref{FIRcomb} (d) at high magnetic field do not
become so close to Kohn modes as in the double dot. The reason is that
electrons localize into two decoupled double dots rather than into
single QDs as in the double dot spectrum of Fig.~\ref{FIRcomb} (b).

All QDM confinements show anticrossings in $\omega_+$. The clearest
anticrossings are seen at low magnetic field strengths. At higher $B$
there are still anticrossings but the energy gap is so small that they
are not visible in the spectra. The transition probabilities in the
lower panels, however, reveal anticrossings as the transition
probability of one mode decreases and the other one increases as a
function of magnetic field. The biggest anticrossing gap in energy is
seen with square-symmetric four-minima QDM. This is interesting, since
in our previous studies the square-symmetric four-minima QDM
resembled the most a parabolic QD whereas double dot and rectangular
four-minima QDM showed greater deviations in the ground state
properties and in the low-lying eigenstates~\cite{MeriQDM1_PRB05}.
Obviously this is not true with the excitation spectrum and with the
higher eigenstates.

\subsection{Analysis of interactions in two-electron FIR spectra}
\label{FIR2}

We conclude that different types of deviations appear in the
spectrum when the confinement potential is not perfectly
parabolic. In order to analyze these deviations in more detail, we
plot the full spectra up to $6$ T magnetic field for both singlet
and triplet states with two interacting electrons and also for two
noninteracting electrons for the comparison. We are interested to see
which features in the spectra are due to electron-electron
interactions and which result from the lower symmetry of the
confinement potential. It is also interesting to see if some features
in the spectra can be used to identify the confinement potential. This
would be useful for experiments where the profile of the underlying
confinement potential is not clear.

\begin{figure}
\hfill
\includegraphics*[width=0.32\columnwidth]{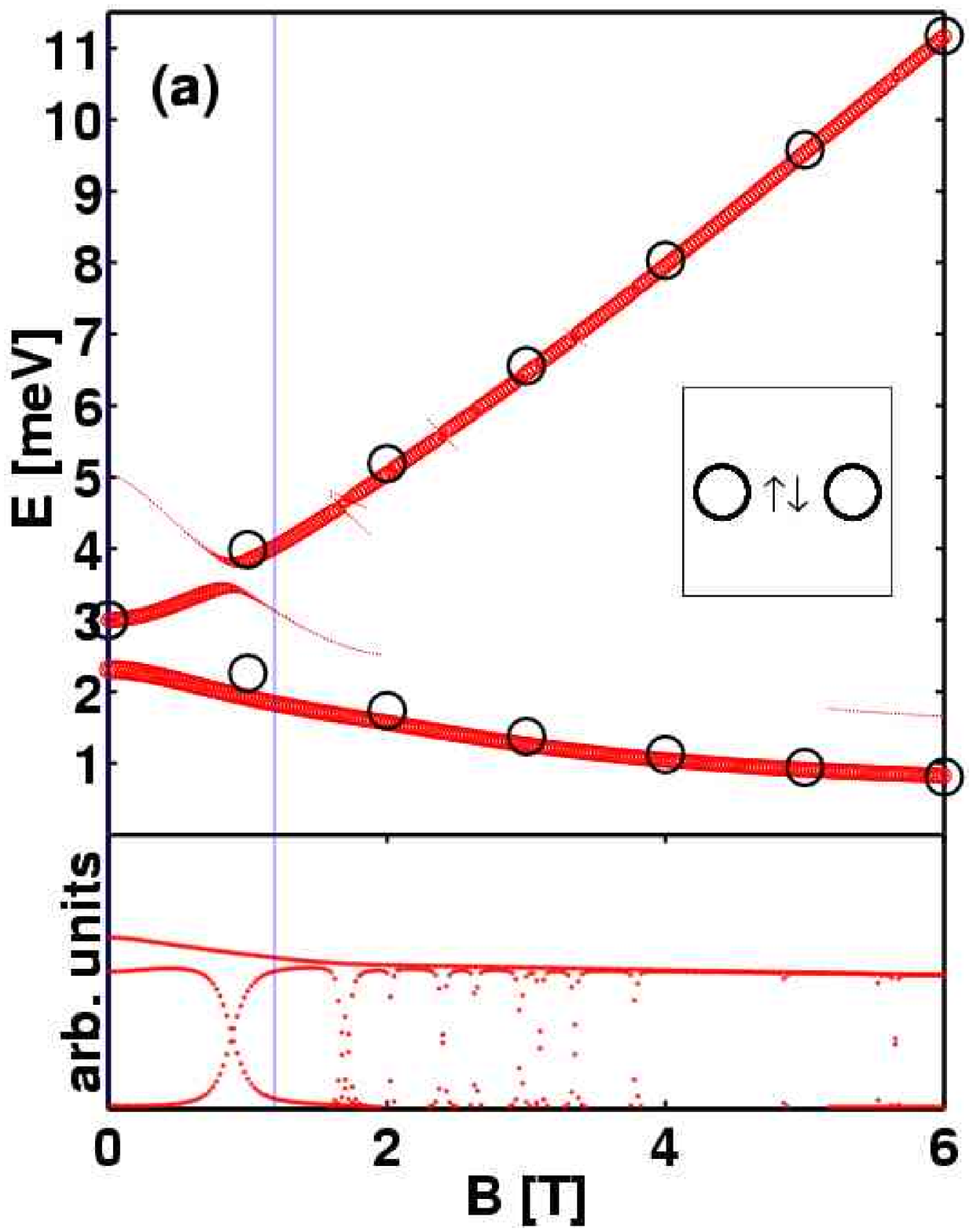}
\includegraphics*[width=0.32\columnwidth]{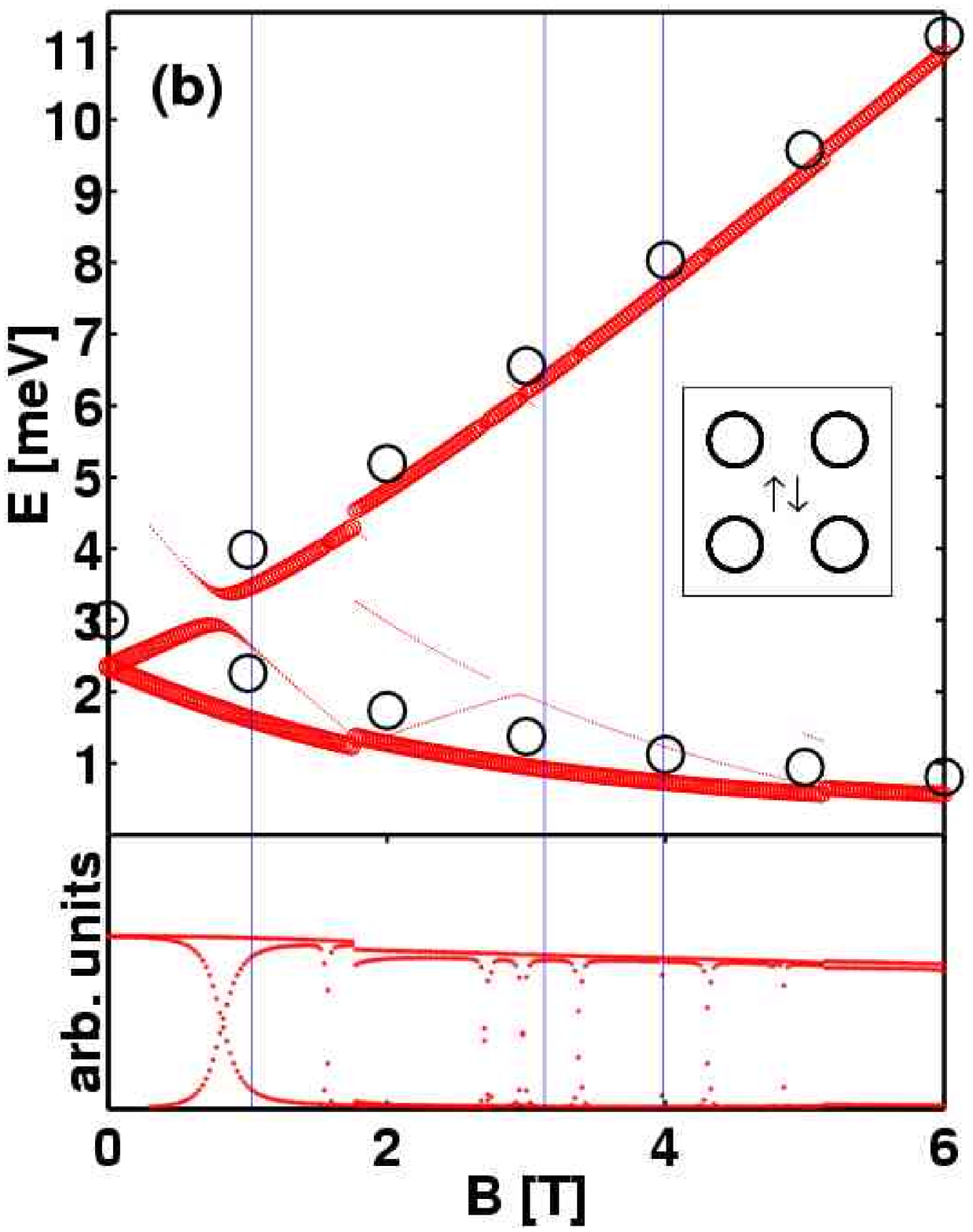}
\includegraphics*[width=0.32\columnwidth]{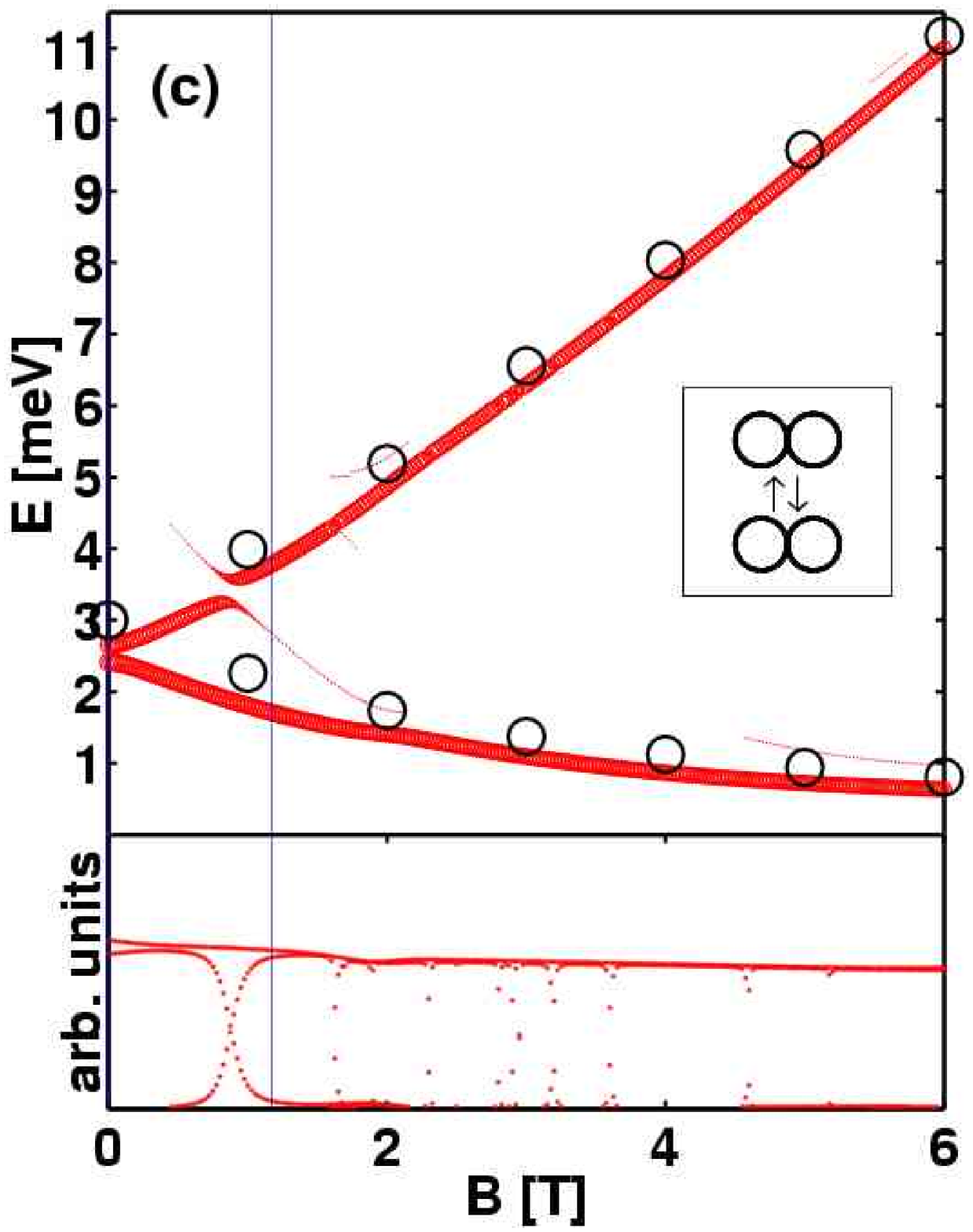}

\hfill
\includegraphics*[width=0.32\columnwidth]{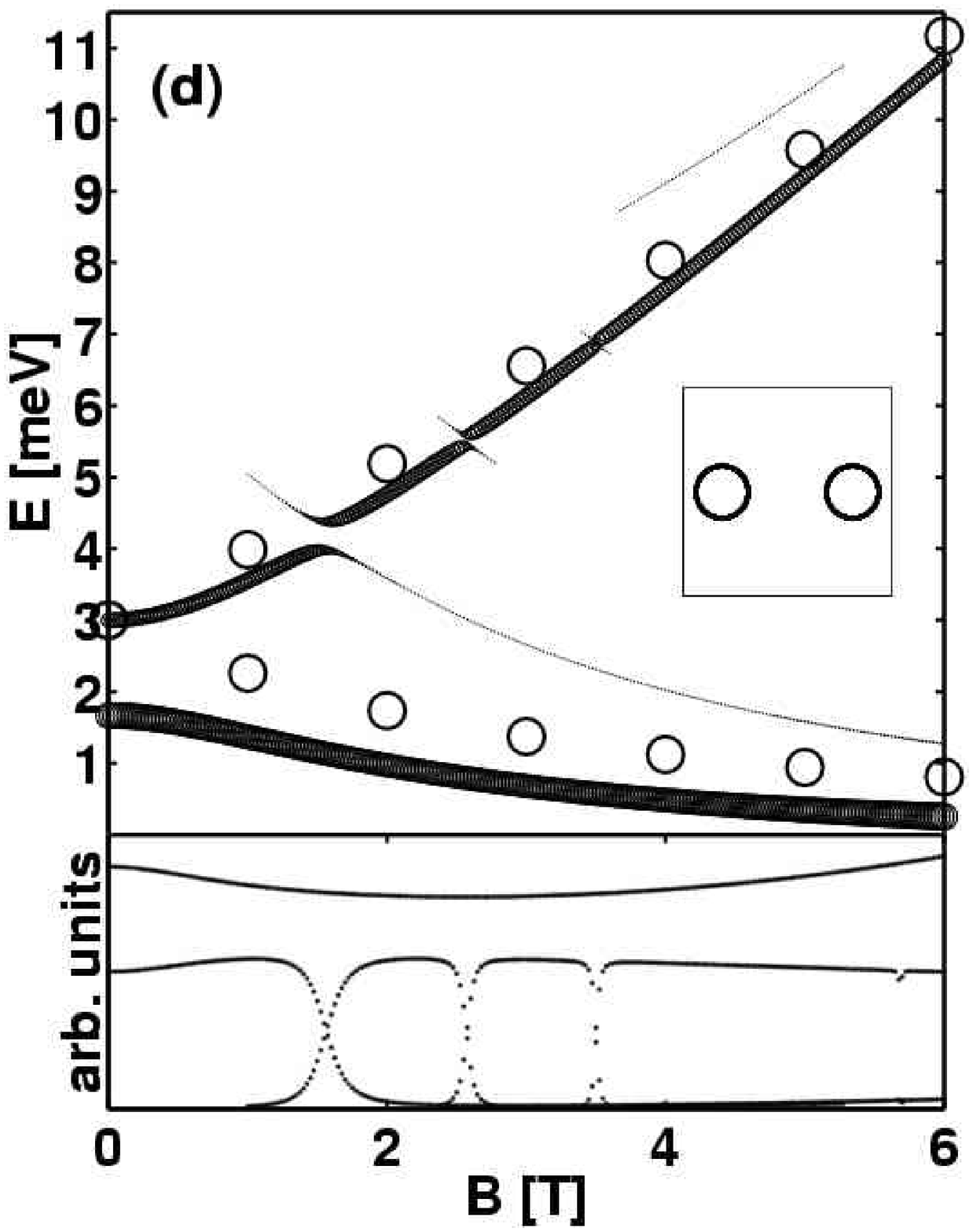}
\includegraphics*[width=0.32\columnwidth]{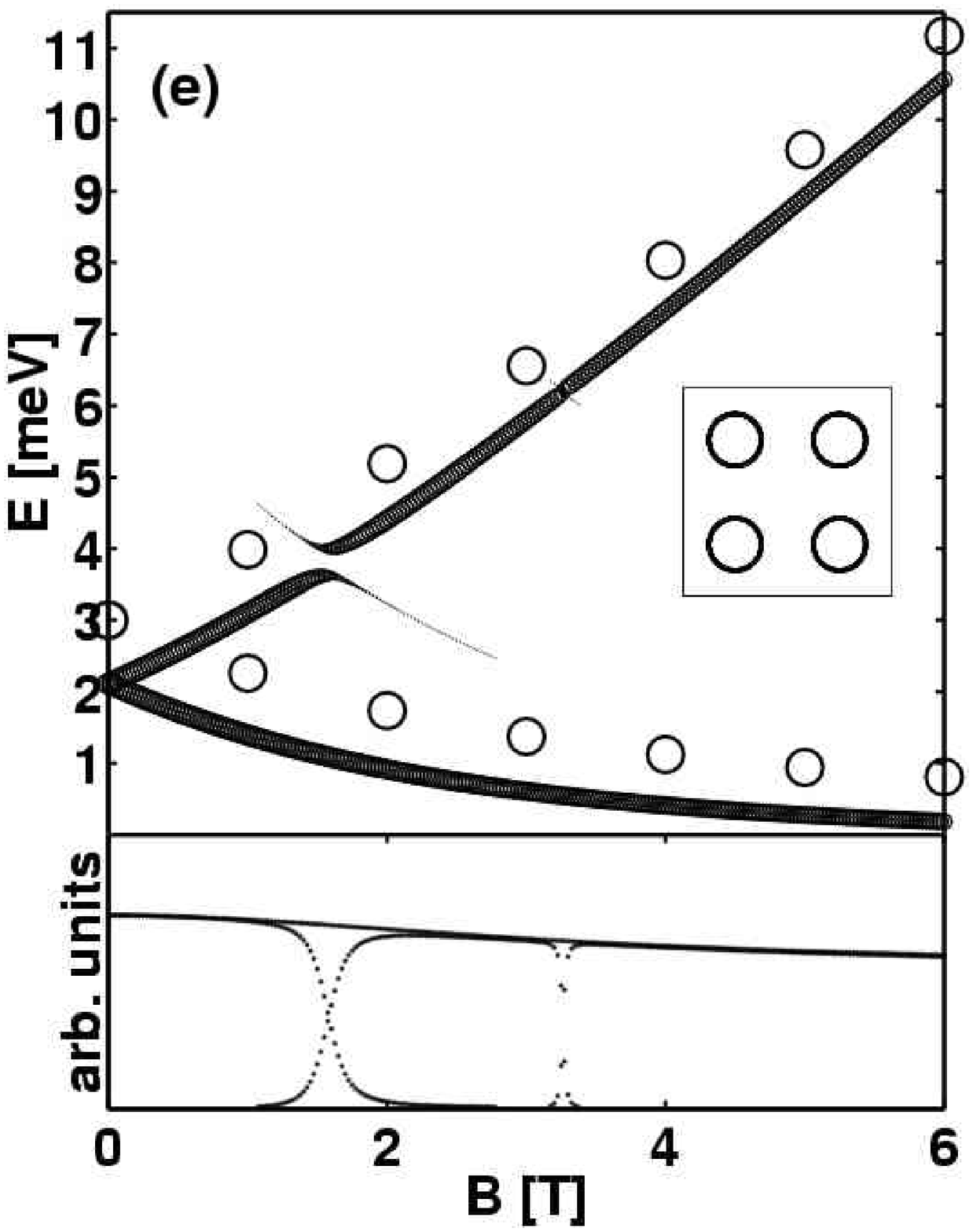}
\includegraphics*[width=0.32\columnwidth]{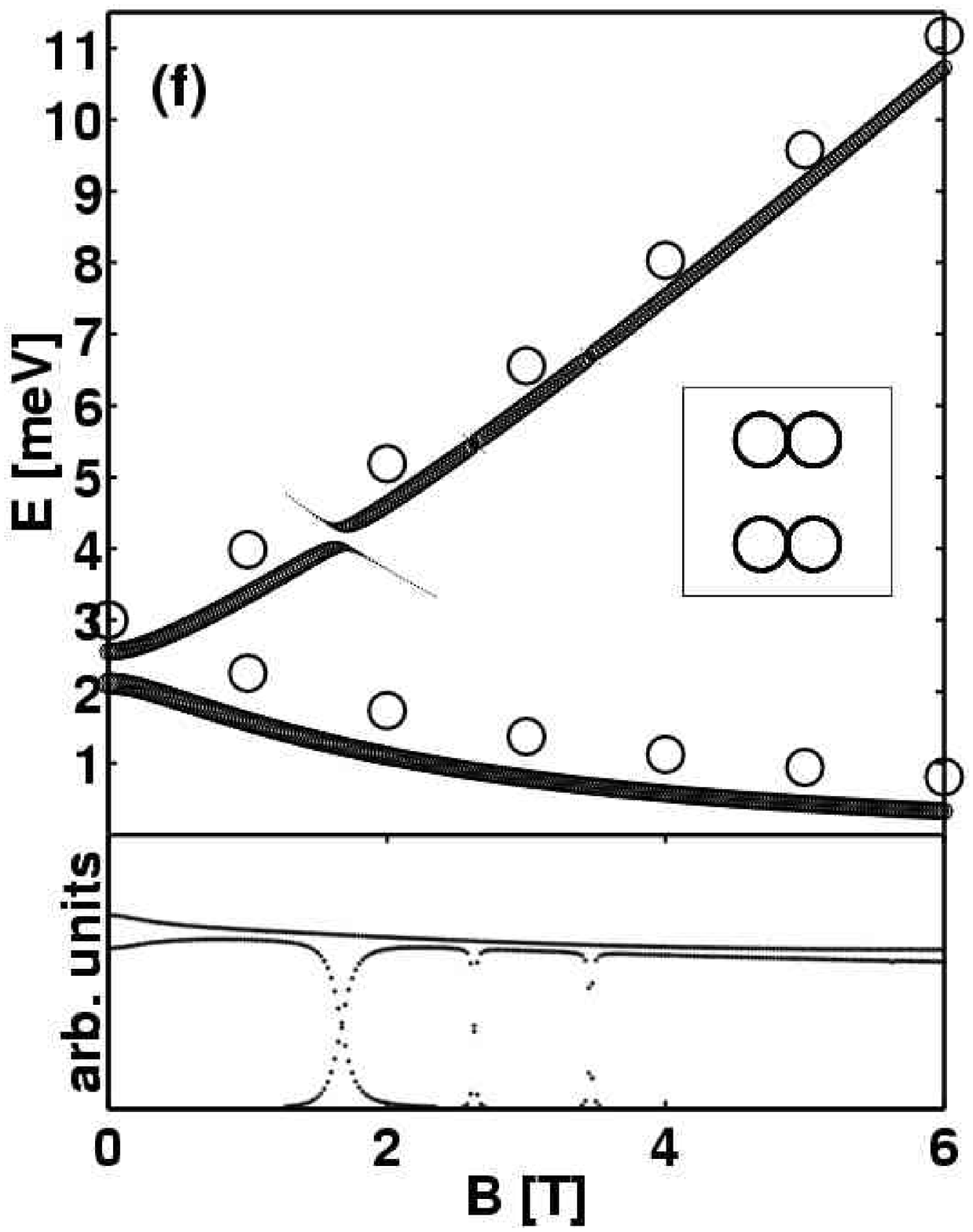}
\caption{Far-infrared spectra of two interacting electrons in the
singlet symmetry in (a)-(c) for double dot, square-symmetric
four-minima quantum-dot molecule, and rectangular-symmetric
four-minima quantum-dot molecule, respectively. (d)-(f) show
corresponding non-interacting far-infrared spectra.}
\label{FIRS0}
\end{figure}

\begin{figure}
\hfill
\includegraphics*[width=0.32\columnwidth]{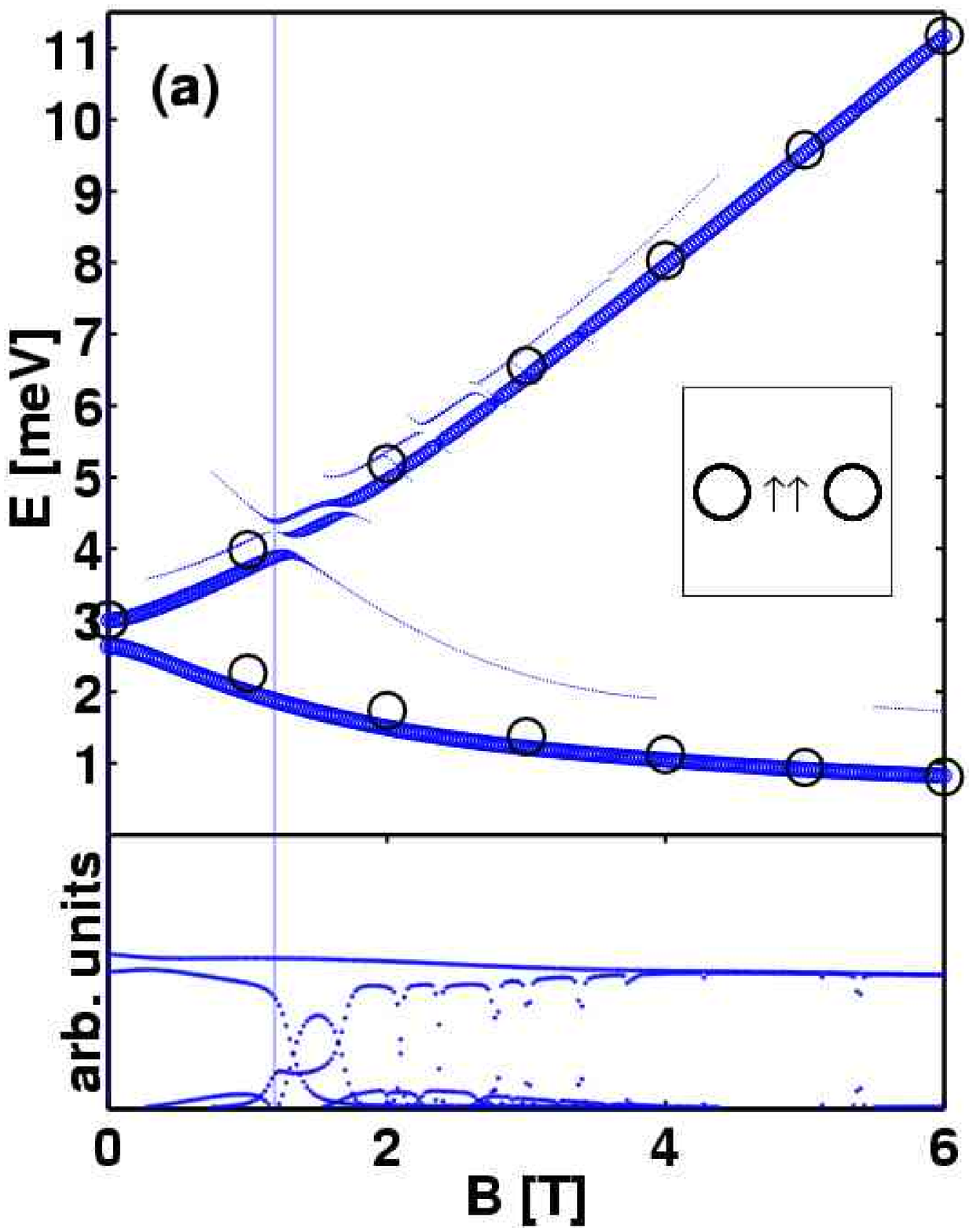}
\includegraphics*[width=0.32\columnwidth]{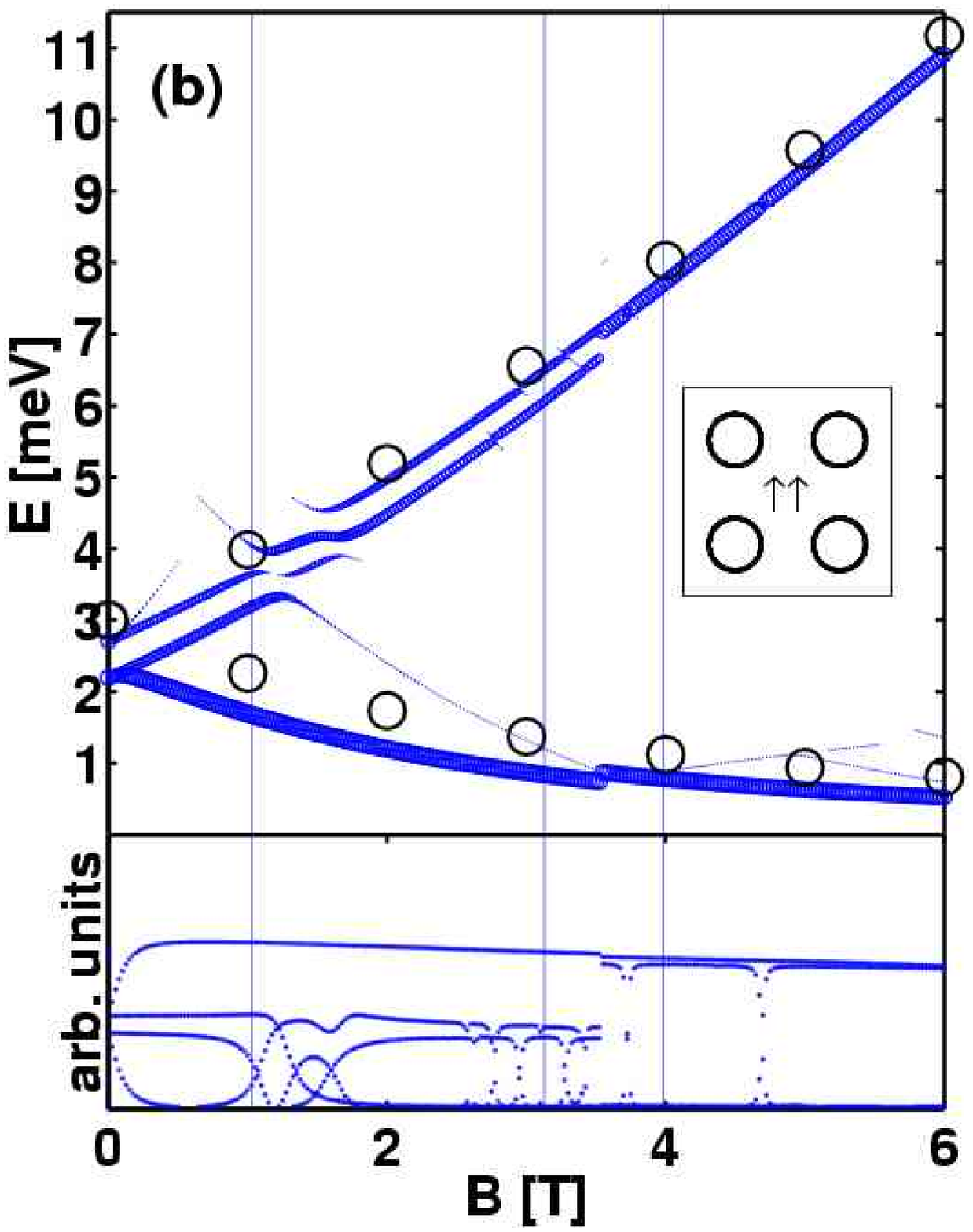}
\includegraphics*[width=0.32\columnwidth]{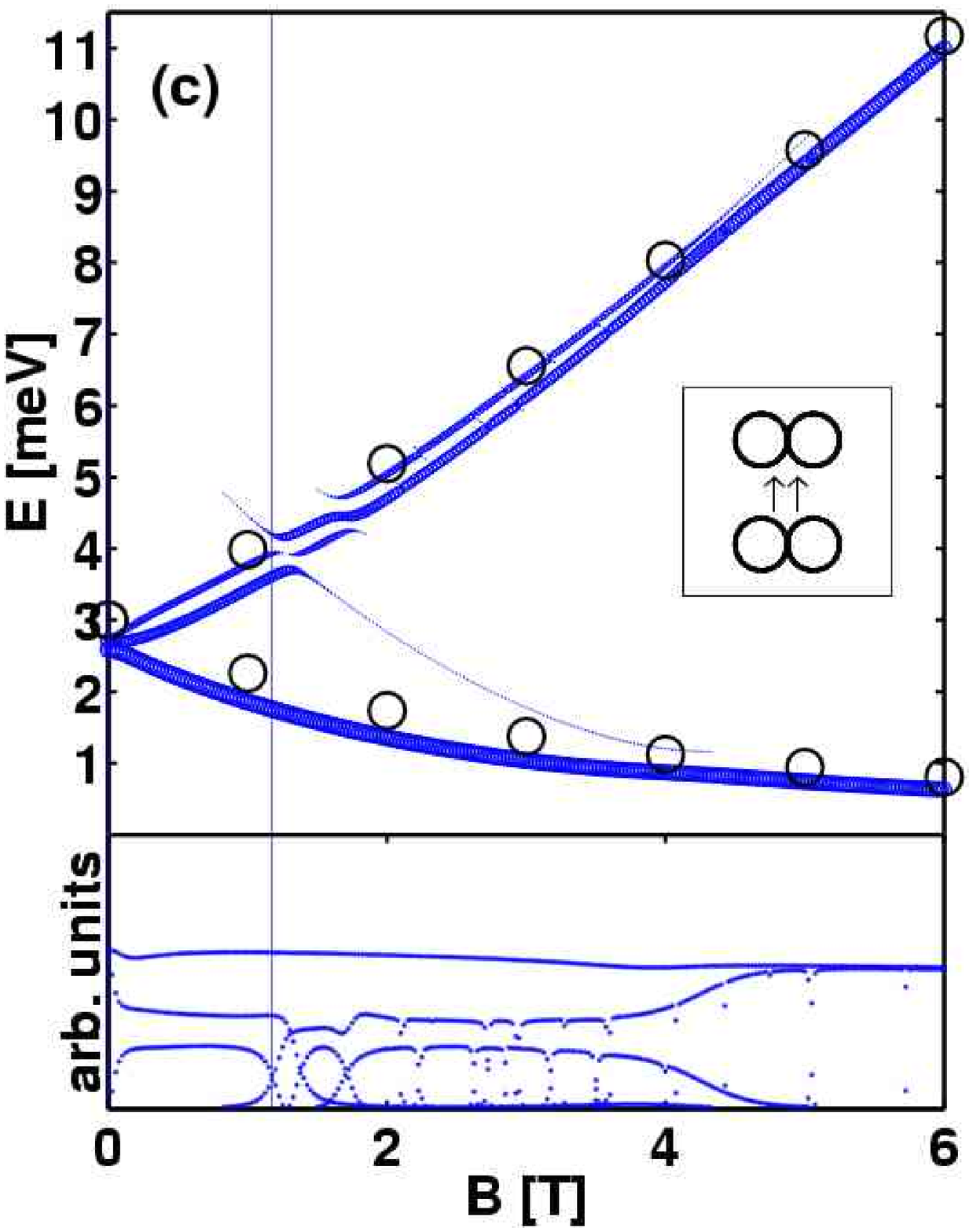}

\hfill
\includegraphics*[width=0.32\columnwidth]{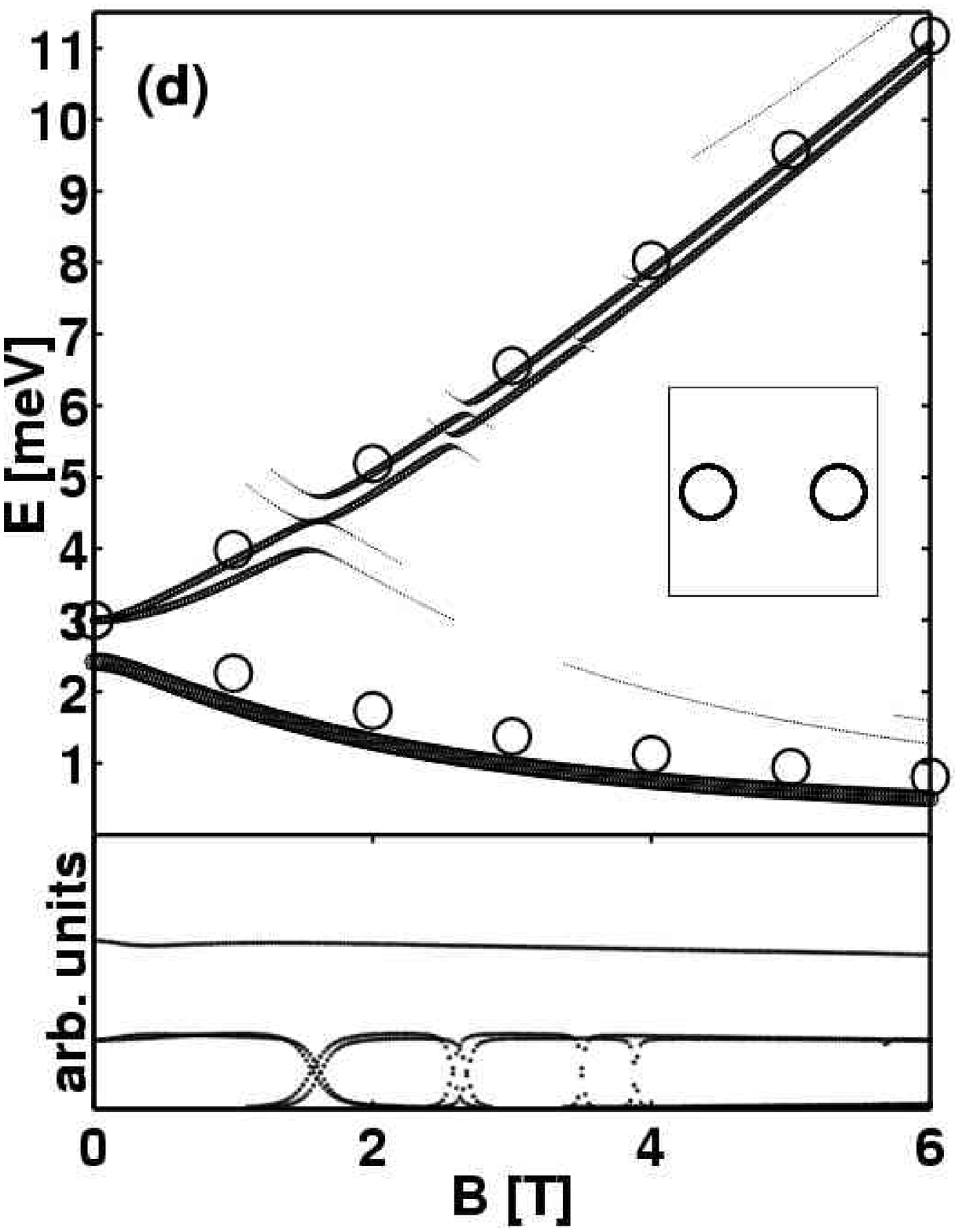}
\includegraphics*[width=0.32\columnwidth]{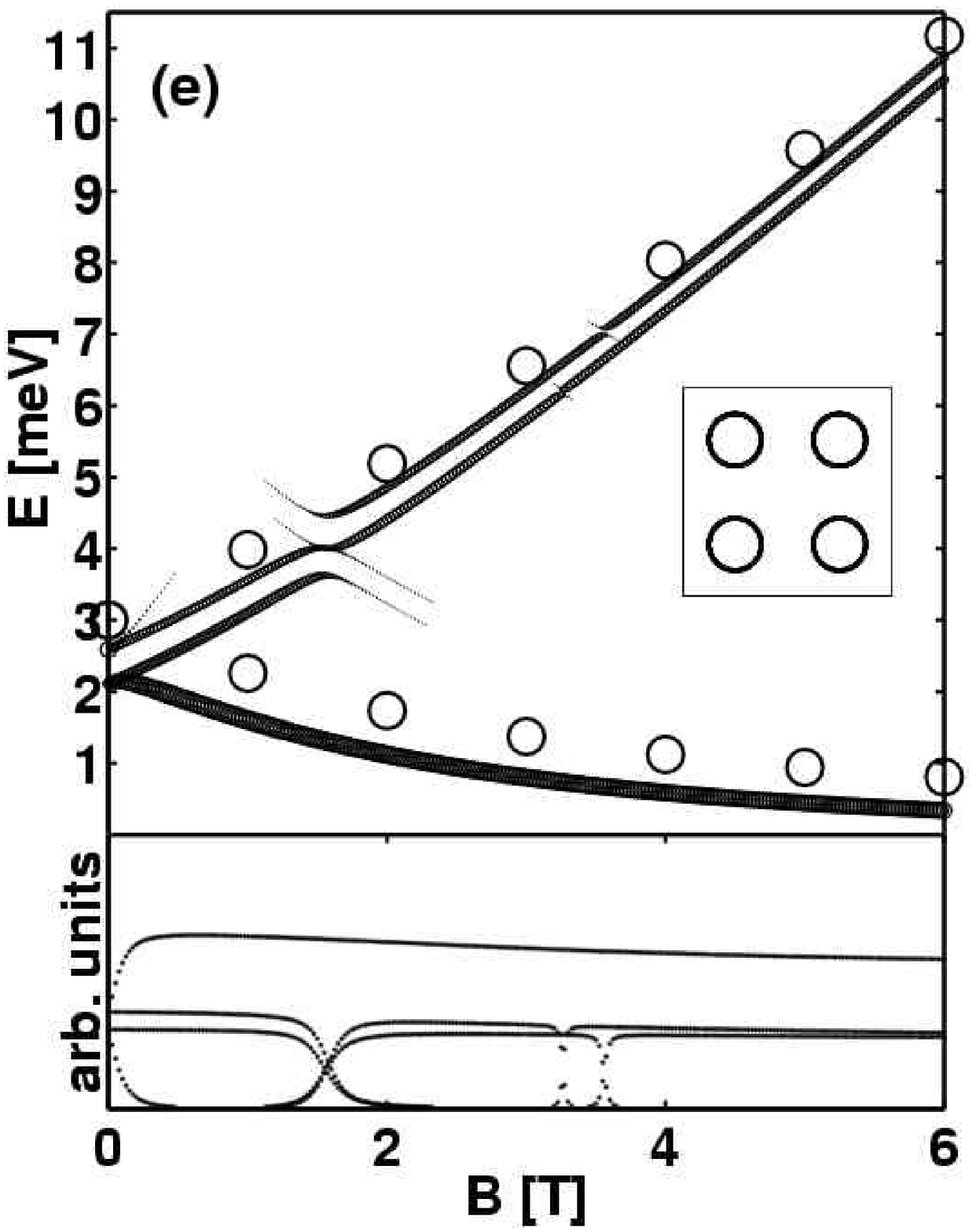}
\includegraphics*[width=0.32\columnwidth]{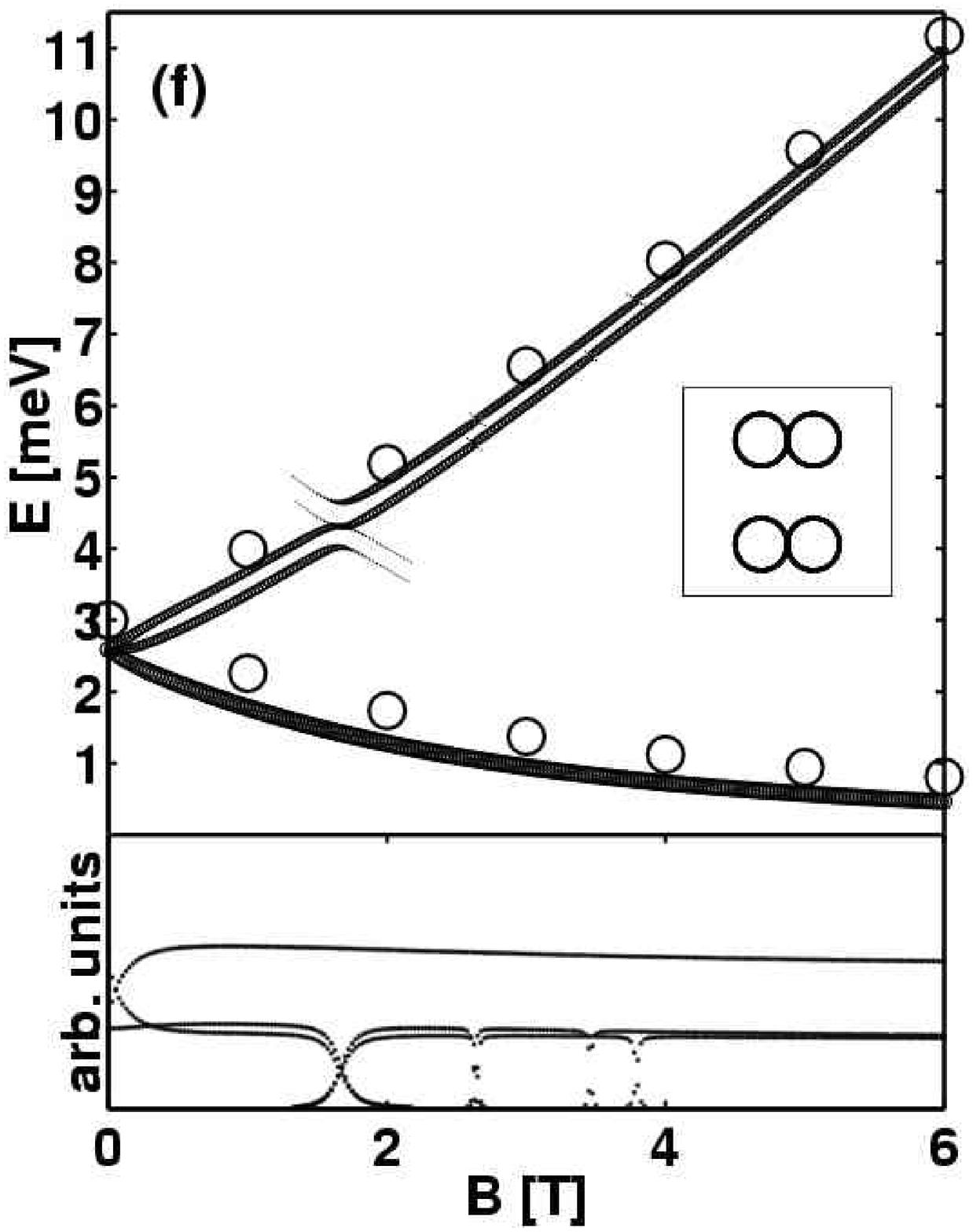}
\caption{Far-infrared spectra of two interacting electrons in the
triplet symmetry in (a)-(c) for double dot, square-symmetric
four-minima quantum-dot molecule, and rectangular-symmetric
four-minima quantum-dot molecule, respectively. (d)-(f) show the
corresponding non-interacting far-infrared spectra.}
\label{FIRS1}
\end{figure}

We will first analyze the two-electron singlet spectra shown in
Fig.~\ref{FIRS0} (a)-(c) for interacting electrons and in (d)-(f) for
noninteracting electrons. The two-electron triplet spectra are shown
in Fig.~\ref{FIRS1} (a)-(c) and the corresponding noninteracting
spectra in (d)-(f). In a real two-electron QDM system, singlet-triplet
transitions must be taken into account and therefore experimentally
observable spectra for two interacting electrons are shown in
Fig.~\ref{FIRcomb}. As the non-interacting singlet spectra, in
Fig.~\ref{FIRS0} (d)-(f), are the same as the single-particle spectra,
these would correspond to experimental FIR spectra with one electron
in the QDM. The non-interacting triplet, on the other hand,
corresponds the FIR spectra of two occupied single-particle levels.


The double dot singlet spectrum in Fig.~\ref{FIRS0} (a) looks
qualitatively very similar to the noninteracting spectrum in
Fig.~\ref{FIRS0} (d). The first difference is that the Kohn modes lie
at the lower energy in the non-interacting spectra compared to the
interacting spectra. The only exception is in the upper mode at $B=0$
where both have excitation energy of $\hbar \omega_0 = 3$ meV. In zero
magnetic field this corresponds to linearly polarized excitation along
$y$ axis where the potential is parabolic and therefore excitations
are not affected by the electron-electron
interactions~\cite{MeriPRL}. Another difference is the zero-field gap
between Kohn modes which is clearly greater in the non-interacting
case.  Higher energy excitations in the interacting case can be
explained with the Coulomb repulsion between the electrons. Coulomb
repulsion effectively steepens the confinement resulting in higher
excitation energies. The anticrossing points are also affected by
interactions whereas the gap remains almost the same.  The
anticrossings can be seen at lower magnetic field values in the
interacting spectrum. Qualitatively the two spectra look very similar
and we conclude that the deviations observed in the spectrum result
from the low symmetry confinement and interactions only shift the
excitation energies and change the anticrossing points.

Generally the same conclusions drawn for the double dot hold for the
square-symmetric four-minima QDM singlet spectra in Fig.~\ref{FIRS0}
(b) and (e). Now in the square-symmetric four-minima QDM the zero
field excitation is degenerate as the two perpendicular directions
($x$ and $y$) have identical confinement profiles.  Interactions shift
the zero field excitation to higher energy due to Coulomb repulsion.
The first anticrossing is again seen at lower $B$ in the interacting
spectrum. However, this time the anticrossing gap is somewhat greater
in the interacting case.  As a new feature, compared to the double
dot, there are now discontinuities in the interacting spectrum.  The
discontinuities in (b) are at the transition points when two singlet
states with a different symmetry cross. At $B \approx 1.8$ T the
lowest-energy singlet state changes from no-vortex state to two-vortex
state and the discontinuity at $B \approx 5.2$ T is at the crossing
point of two-vortex and four-vortex states (see
Ref.~\cite{MeriQDM1_PRB05} for more details on vortex states).
However, the discontinuities are seen at the magnetic field values
where the true ground state is triplet and therefore the
discontinuities in Fig.~\ref{FIRS0} (b) would not be observed in FIR
experiments.

In the rectangular-symmetric four-minima QDM (Fig.~\ref{FIRS0} (c) and
(f)) the confinement profile is non-identical and non-parabolic in
both perpendicular directions. Therefore both zero-field excitations
are lower than $3$ meV and there is also a gap between the two
modes. Again excitation energies as a function of magnetic field are
higher in the interacting spectrum and anticrossings are seen at the
lower magnetic field values compared to the noninteracting
spectrum. Interesting point, when comparing interacting and
noninteracting spectra, is that the zero-field excitation energy of
the lower mode shifts to $0.30$ meV higher and the higher mode shifts
only $0.09$ meV higher when the interactions are turned on. It means
that the Coulomb repulsion is over three times more significant in the
excitation along the long axis ($y$ axis) than it is along the short
axis ($x$ axis). Electrons are localized into two distant double
dots. Coulomb repulsion is more important between the two double dots
as it is inside one double dot, {\it i.e.}, along long and short axes.


Next, we do similar comparisons for the triplet spectra of QDMs.  The
interacting spectra are shown in Fig.~\ref{FIRS1} (a)-(c) and
non-interacting in Fig.~\ref{FIRS1} (d)-(f). We will first compare
double dot interacting triplet spectra of Fig.~\ref{FIRS1} (a) to the
noninteracting spectra of Fig.~\ref{FIRS1} (d). In general, the same
analysis applies to triplet spectra as for the singlet spectra of
double dot. However, in the triplet state there is an additional mode,
$\omega_{+2}$, above the main branch, $\omega_+$, in
Fig.~\ref{FIRS1}(a). The additional mode has clearly weaker transition
probability than the $\omega_+$ mode. In the noninteracting case the
upper mode is split to two modes which both have equal transition
probabilities. In the interacting spectrum the upper mode has lower
transition probability and soon after $4$ T it completely vanishes
from the spectrum.

In the square-symmetric four-minima QDM also the interacting triplet
spectrum, as well as the non-interacting spectrum, has two upper modes
with almost equal transition probabilities (Fig.~\ref{FIRS1} (b)). In
four-minima QDM the split-off branch suddenly changes to one mode at
$B \approx 3.5$ T. There the triplet ground state changes from the
one-vortex to three-vortex solution (see Ref.~\cite{MeriQDM1_PRB05}
for more details on vortex states). Now the two triplet states
actually cross and there occurs a discontinuity in the FIR
spectrum. Yet as the singlet is the true ground state of the
square-symmetric four-minima QDM between $B\approx3$ and $4$ T, the
discontinuity of Fig.~\ref{FIRS1} (b) cannot be seen in experiments.
In triplet spectra there are much smaller differences between the
excitation energies of interacting and noninteracting electrons
compared to the singlet spectra. In the singlet state electrons of
opposite spins can occupy the same single-particle levels in the
many-body configurations and therefore Coulomb repulsion is more
significant for the singlet state. In other words, 
the Pauli exclusion principle keeps the same-spin electrons further
apart and therefore Coulomb repulsion has a lesser effect on the
triplet state compared to the singlet state.

In the rectangular-symmetric four-minima QDM, the zero-field gap is
really small and not visible in the interacting spectra of
Fig.~\ref{FIRS1} (c) and neither in the non-interacting triplet
spectra of Fig.~\ref{FIRS1} (f).  The transition probabilities of the
two main branches ($\omega_+$ and $\omega_-$), in the interacting
spectrum, are almost equal at zero field, but soon after the magnetic
field strength increases the transition probability of $\omega_{+}$
decreases at the same time as the transition probability of
$\omega_{+2}$ increases. After $B=4$ T the $\omega_{+2}$ starts to
weaken again as the symmetry of the triplet ground state is changing.
The $\omega_{+2}$ dies out continuously.  The two upper modes of the
non-interacting spectrum have almost equal transition probabilities.
The interacting modes are higher in energy, also at $B=0$, and
anticrossings shift to lower $B$ in the interacting
spectrum. Anticrossing gaps are not noticeably affected by the
interactions.

\section{FIR spectra of noninteracting electrons}
\label{FIR3}

A general conclusion drawn above is that the noninteracting spectra
show more deviations from the Kohn modes than the interacting spectra.
All the same features are present in noninteracting and interacting
spectra. In addition, the non-interacting spectra show in all cases
similar structure as the interacting spectra. Interactions only shift
some of the features seen in the spectra. Therefore we calculate FIR
spectra of non-interacting electrons up to $N=6$ electrons in the QDMs
to see if more electrons, even if non-interacting, produce changes in
the FIR spectra.

\begin{figure}
\hfill
\includegraphics*[width=0.32\columnwidth]{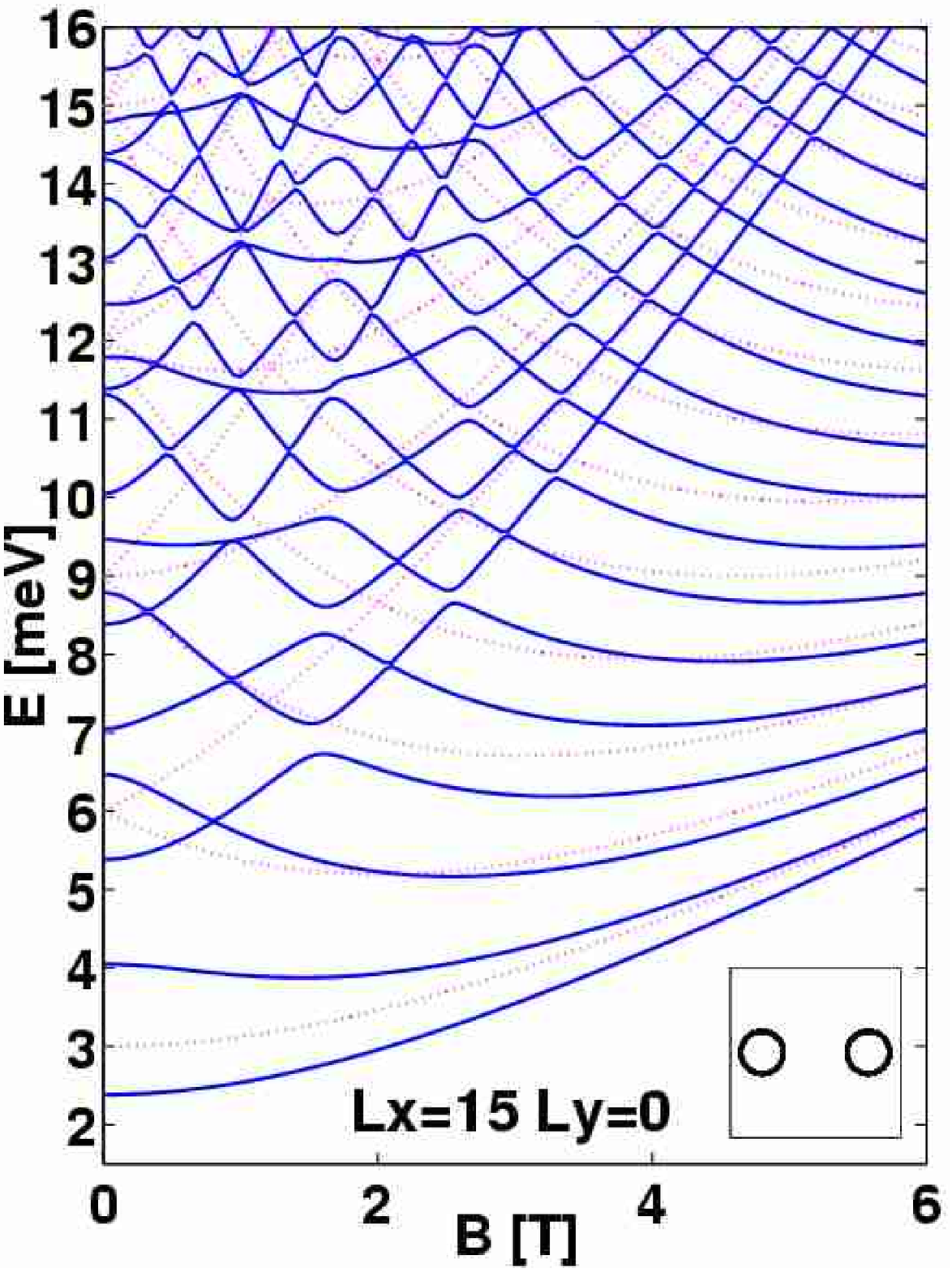}
\includegraphics*[width=0.32\columnwidth]{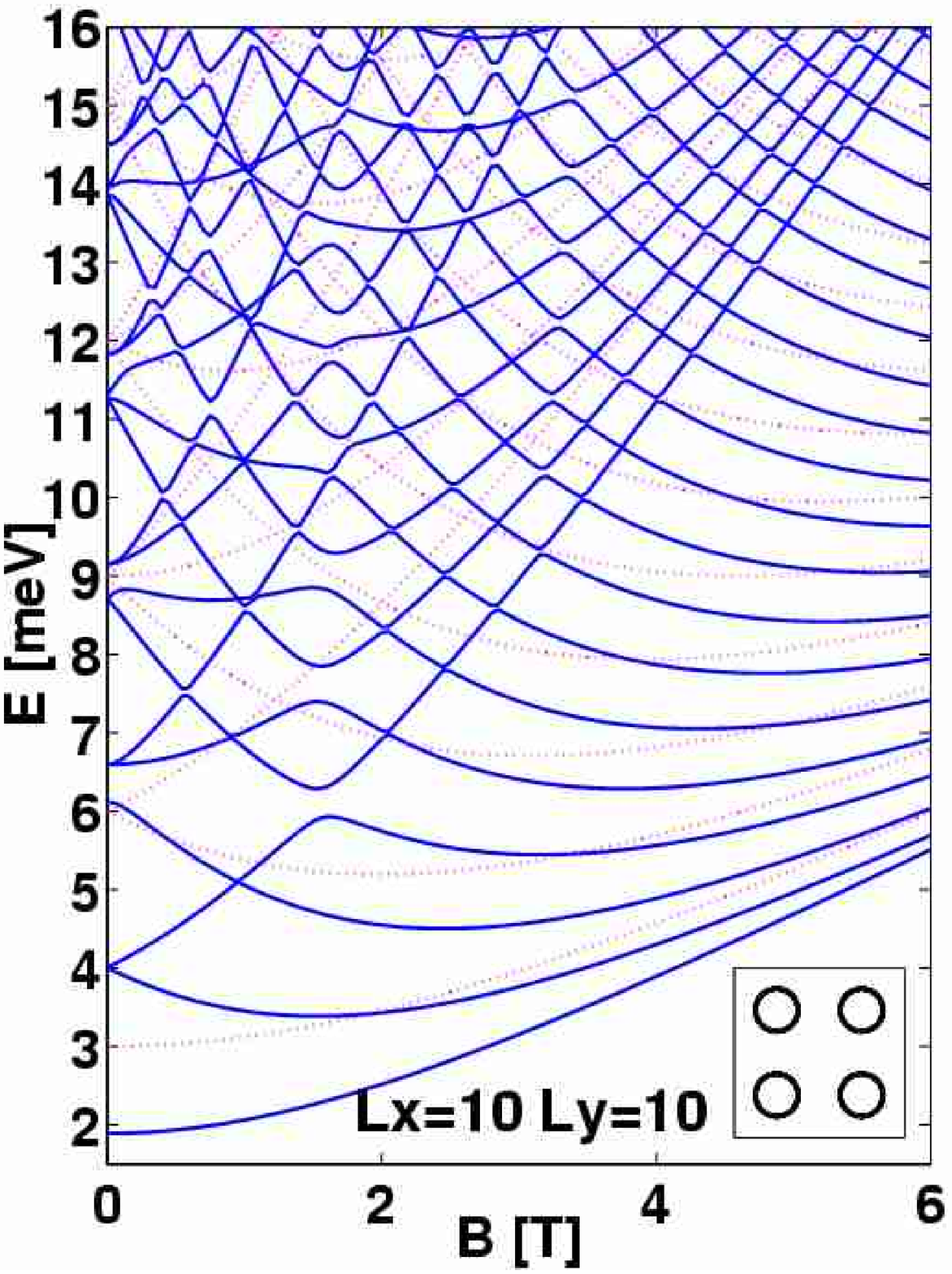}
\includegraphics*[width=0.32\columnwidth]{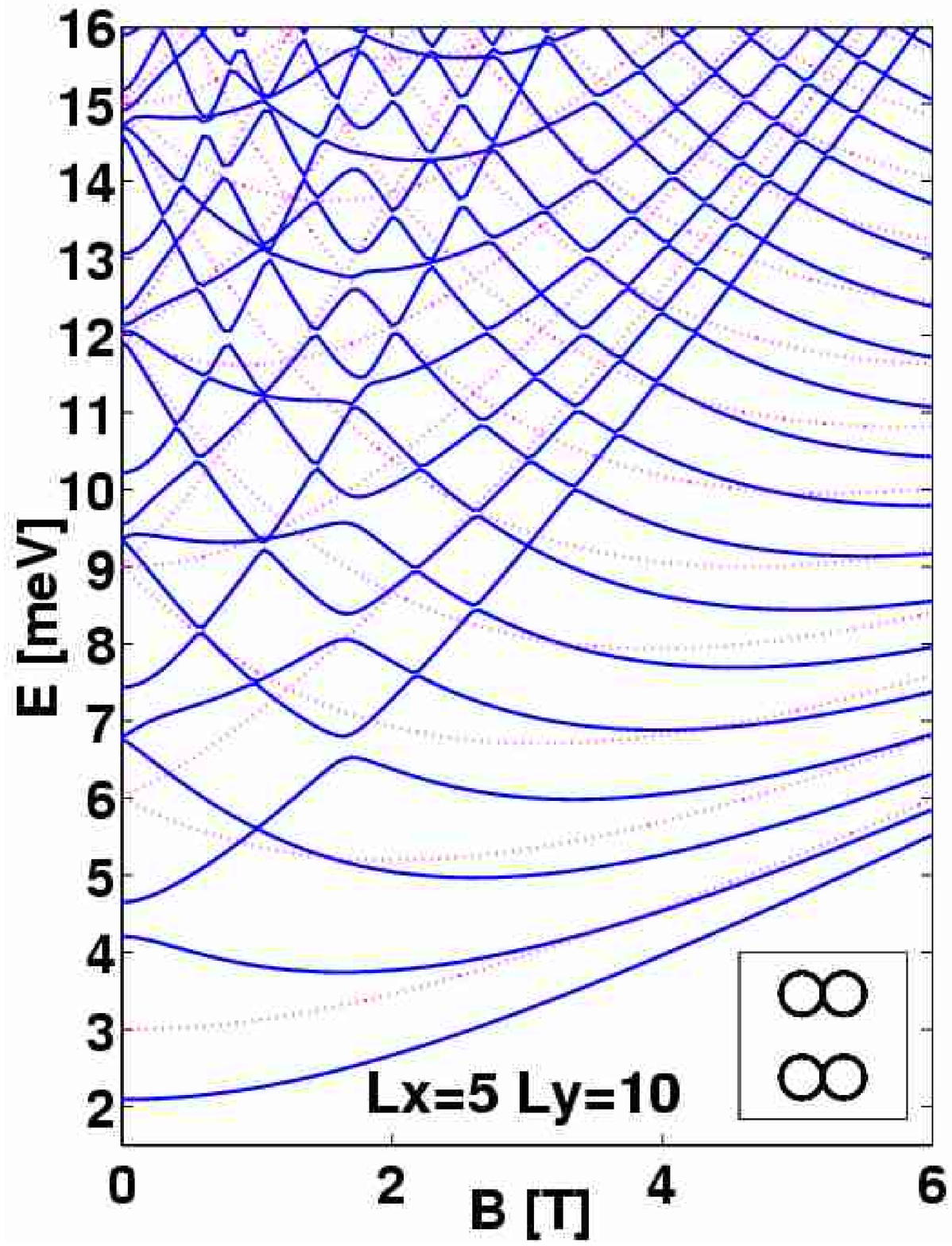}
\caption{Single-particle energy levels of double dot ($L_x=15,L_y=0$
  nm), four-minima square-symmetric quantum-dot molecule ($L_x=L_y=10$
  nm) and four-minima rectangular symmetric quantum-dot molecule
  ($L_x=5,L_y=10$ nm). Dotted lines show single-particle (Fock-Darwin)
  energy levels of parabolic ($\hbar\omega_0=3$ meV) quantum dot.}
\label{ele_yks}
\end{figure}

Before presenting the FIR spectra we plot single-particle energy
levels of all three QDM confinements in Fig.~\ref{ele_yks}. The solid
lines show QDM energy levels and the dotted lines Fock-Darwin energy
levels of a parabolic QD. In QDMs the energy levels shift to lower
energies compared to parabolic QD. Also many anticrossings and
zero-field splittings of energy levels are visible in QDMs.

\begin{figure}
\hfill
\includegraphics*[width=0.32\columnwidth]{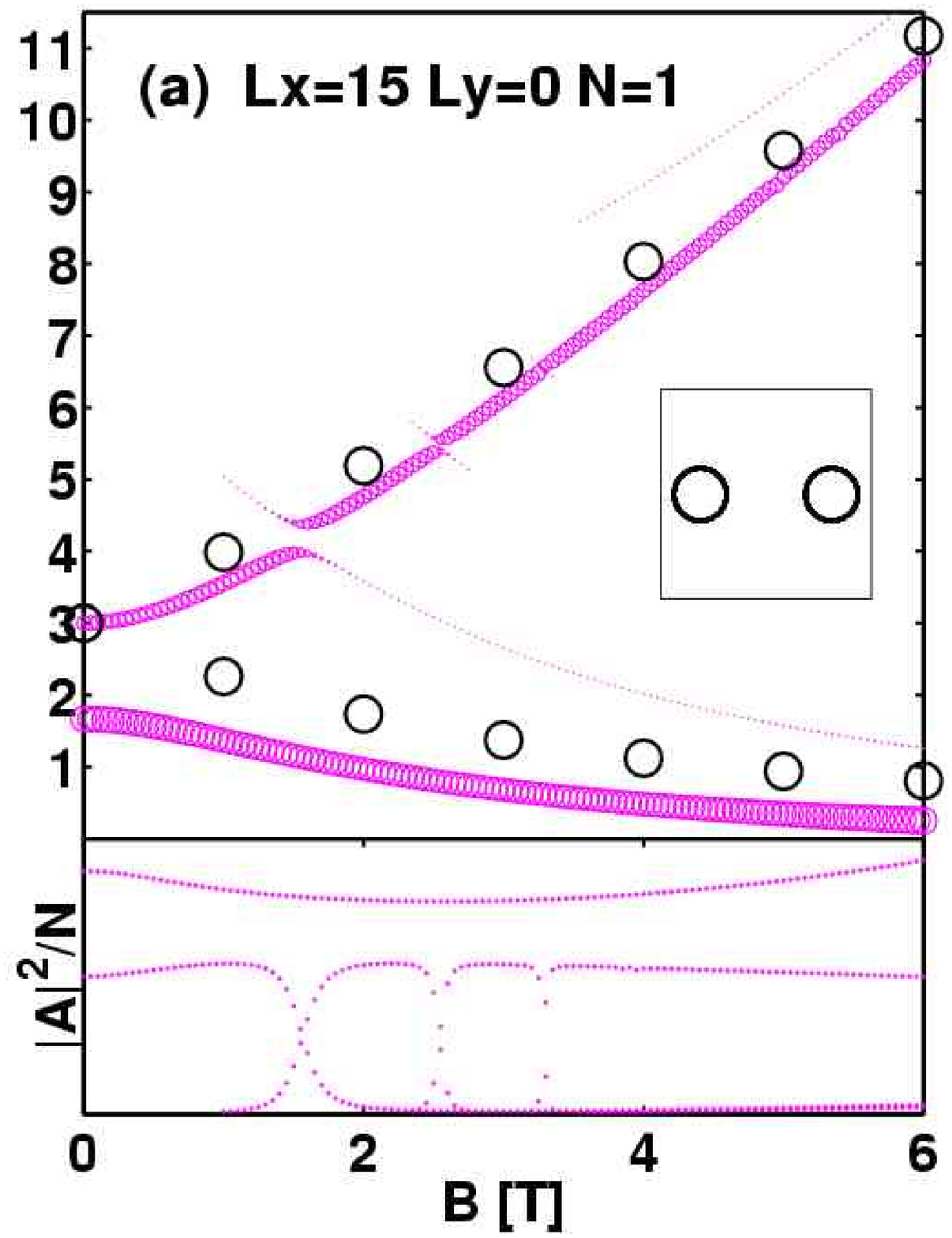}
\includegraphics*[width=0.32\columnwidth]{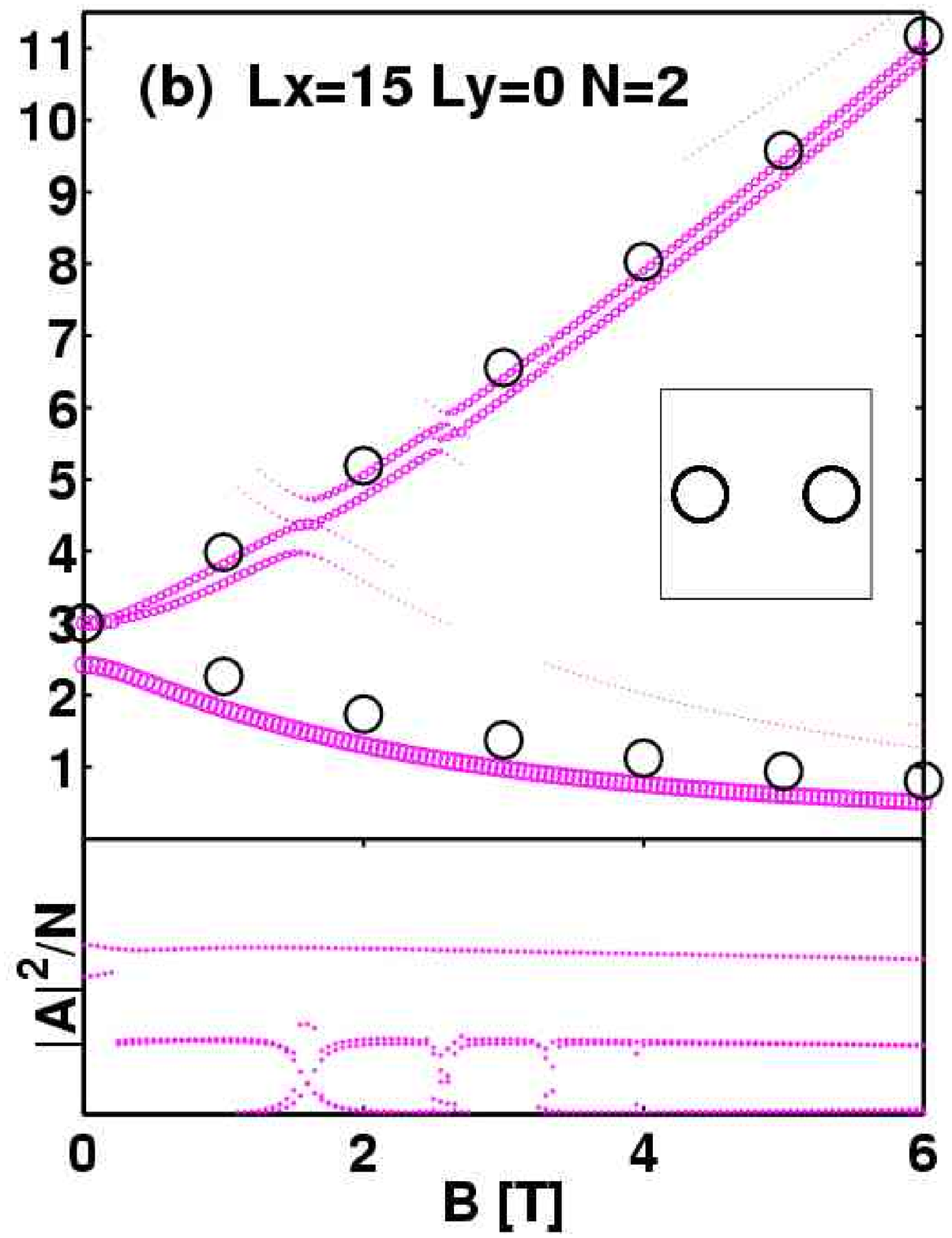}
\includegraphics*[width=0.32\columnwidth]{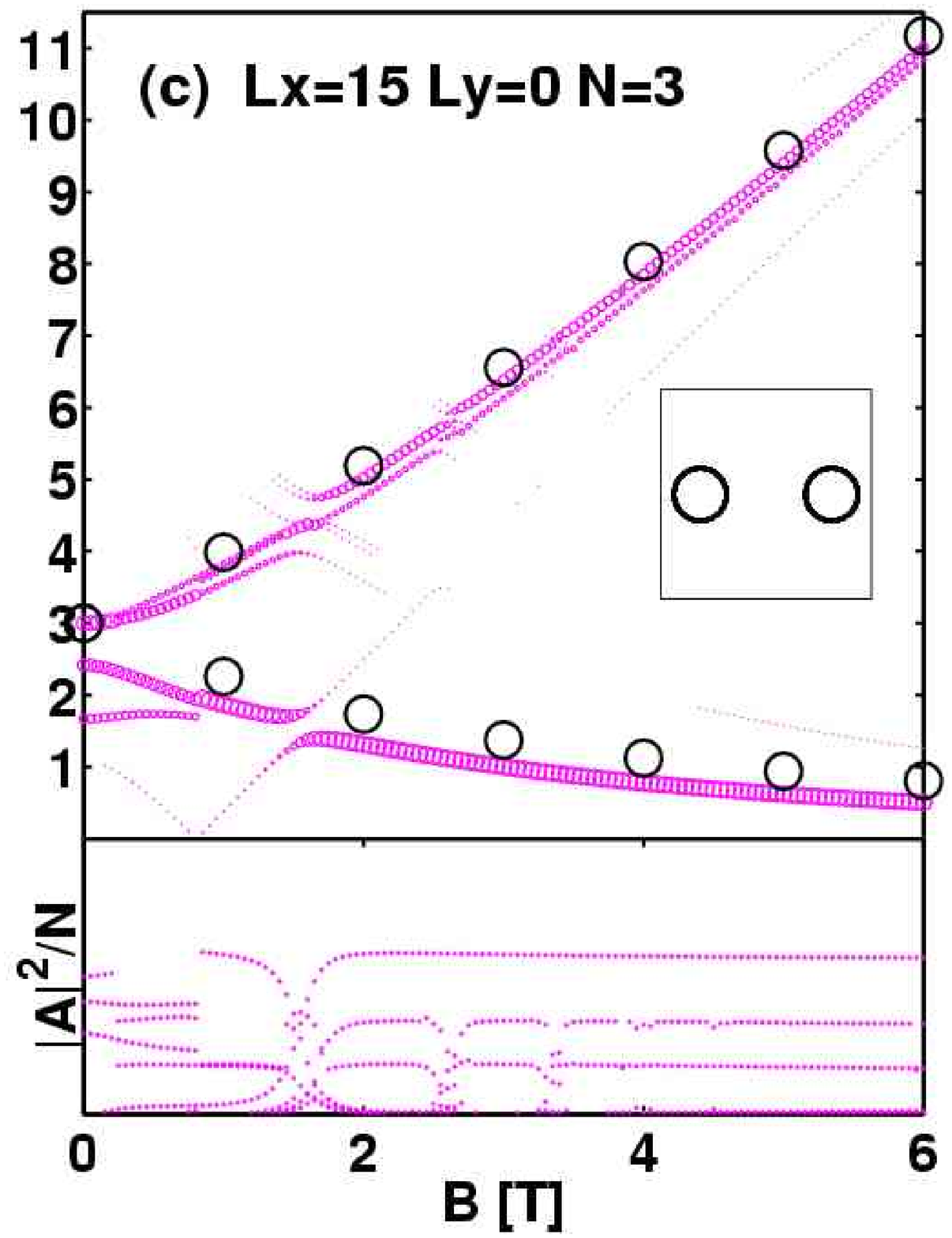}

\hfill
\includegraphics*[width=0.32\columnwidth]{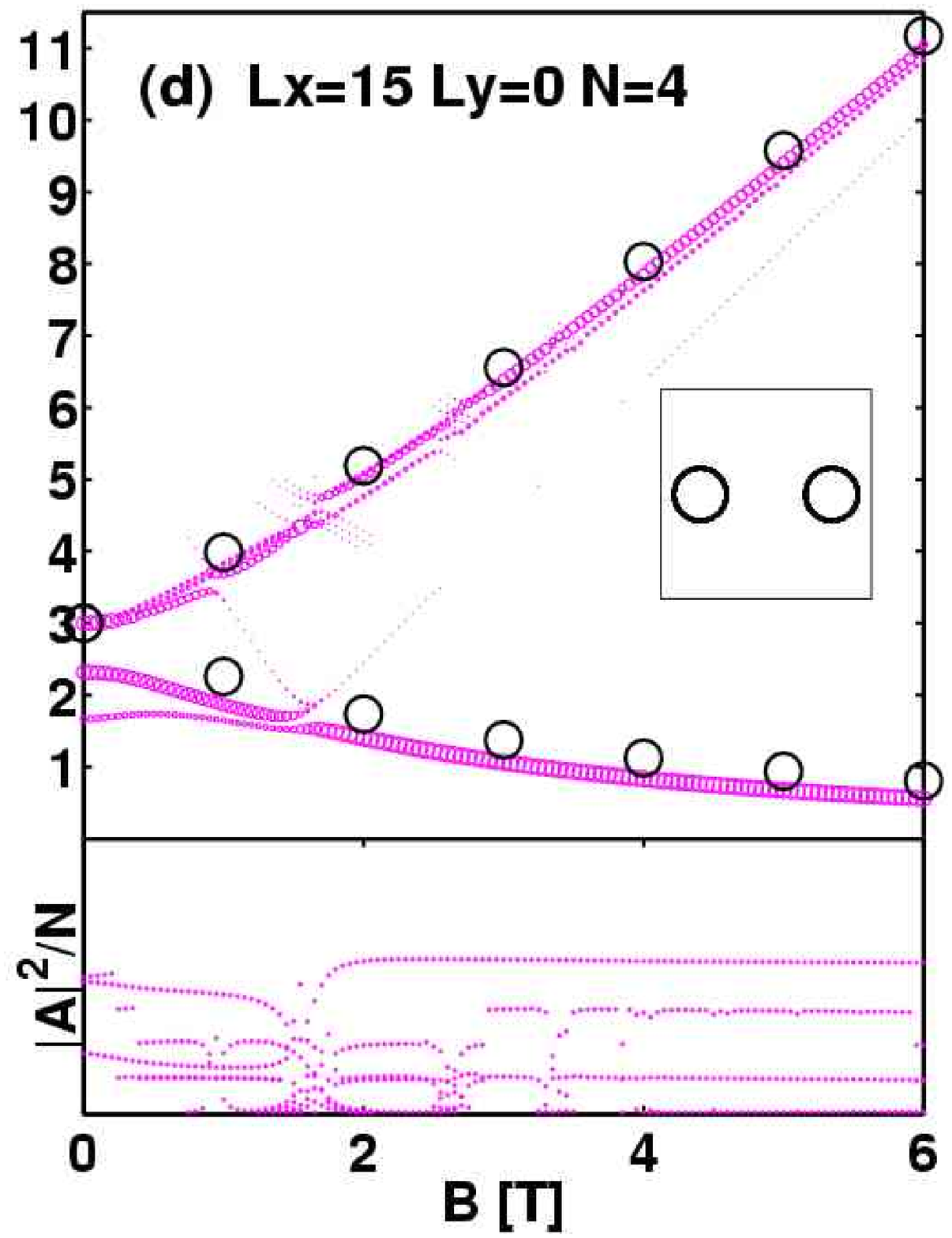}
\includegraphics*[width=0.32\columnwidth]{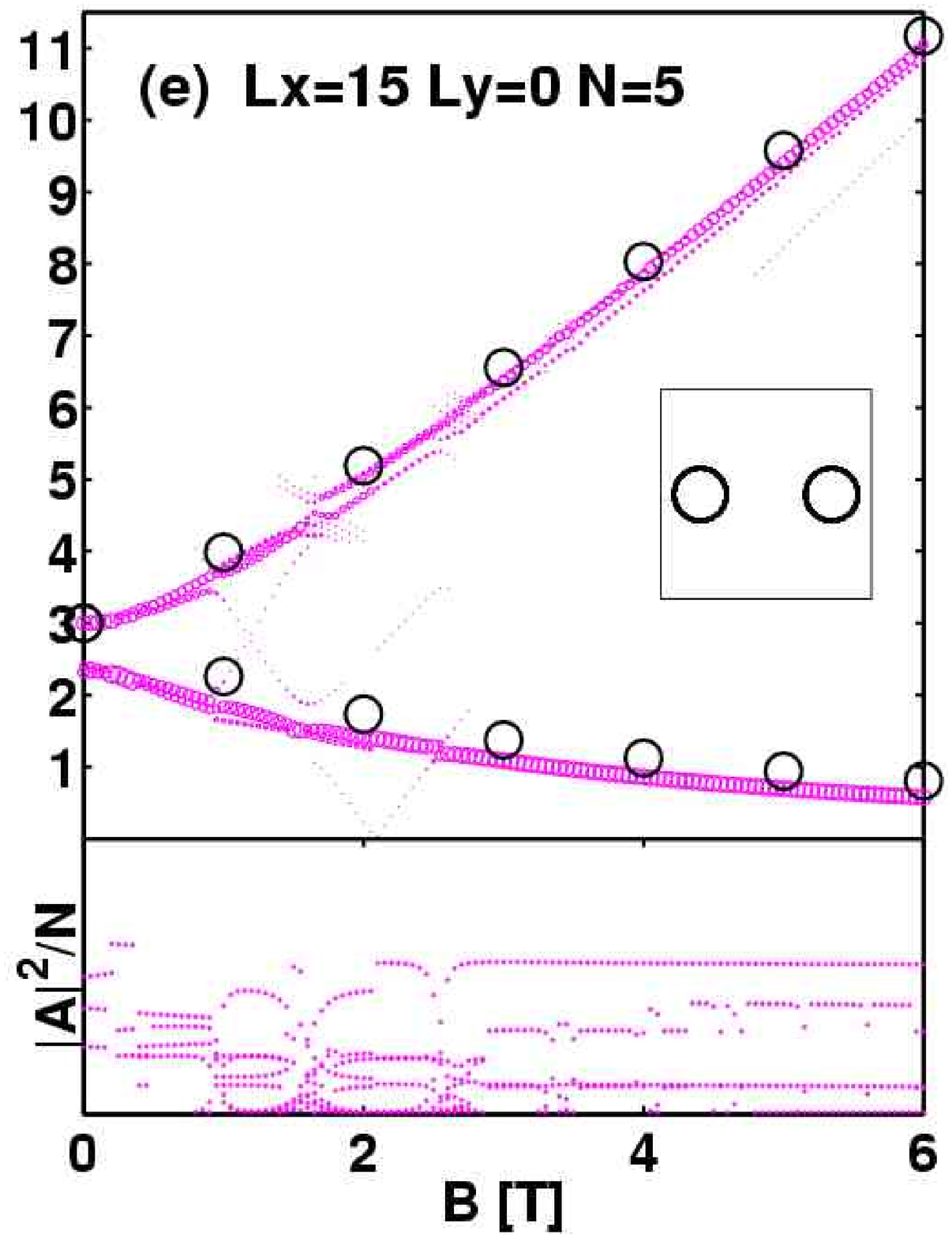}
\includegraphics*[width=0.32\columnwidth]{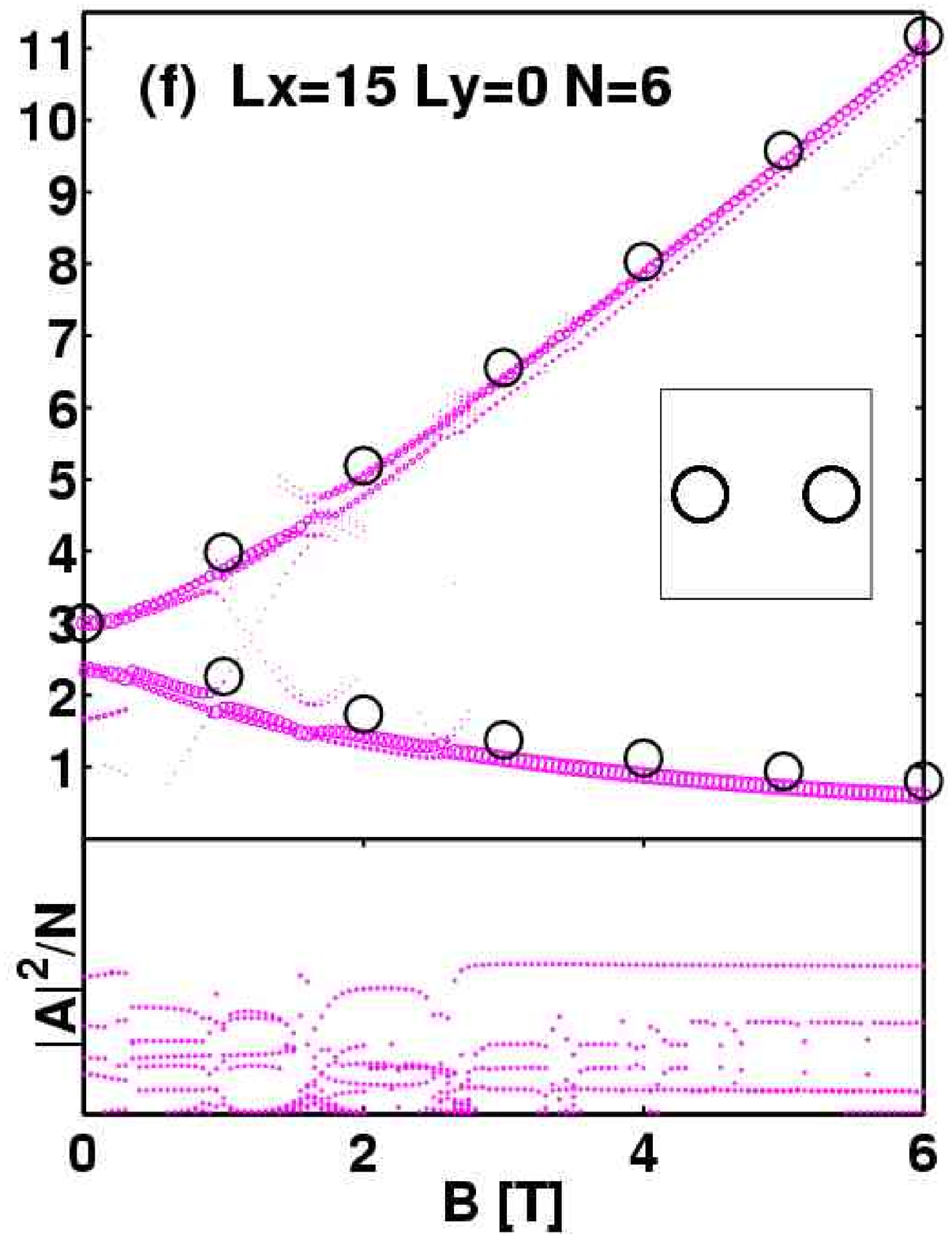}
\caption{Far-infrared spectra for $N=1-6$ non-interacting electrons in
double dot ($L_x=15,L_y=0$ nm).}
\label{YKSLx15Ly0}
\end{figure}

\begin{figure}
\hfill
\includegraphics*[width=0.32\columnwidth]{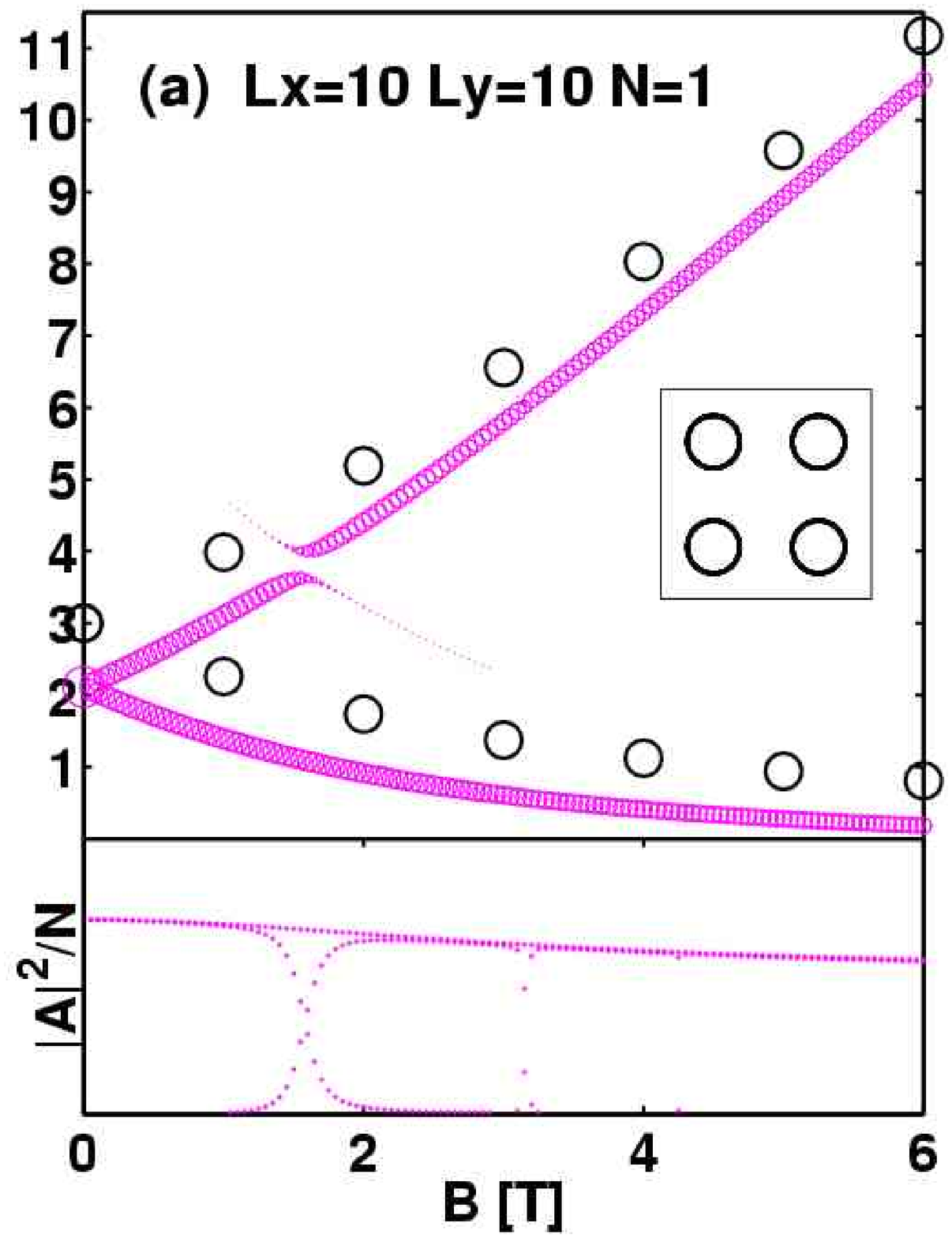}
\includegraphics*[width=0.32\columnwidth]{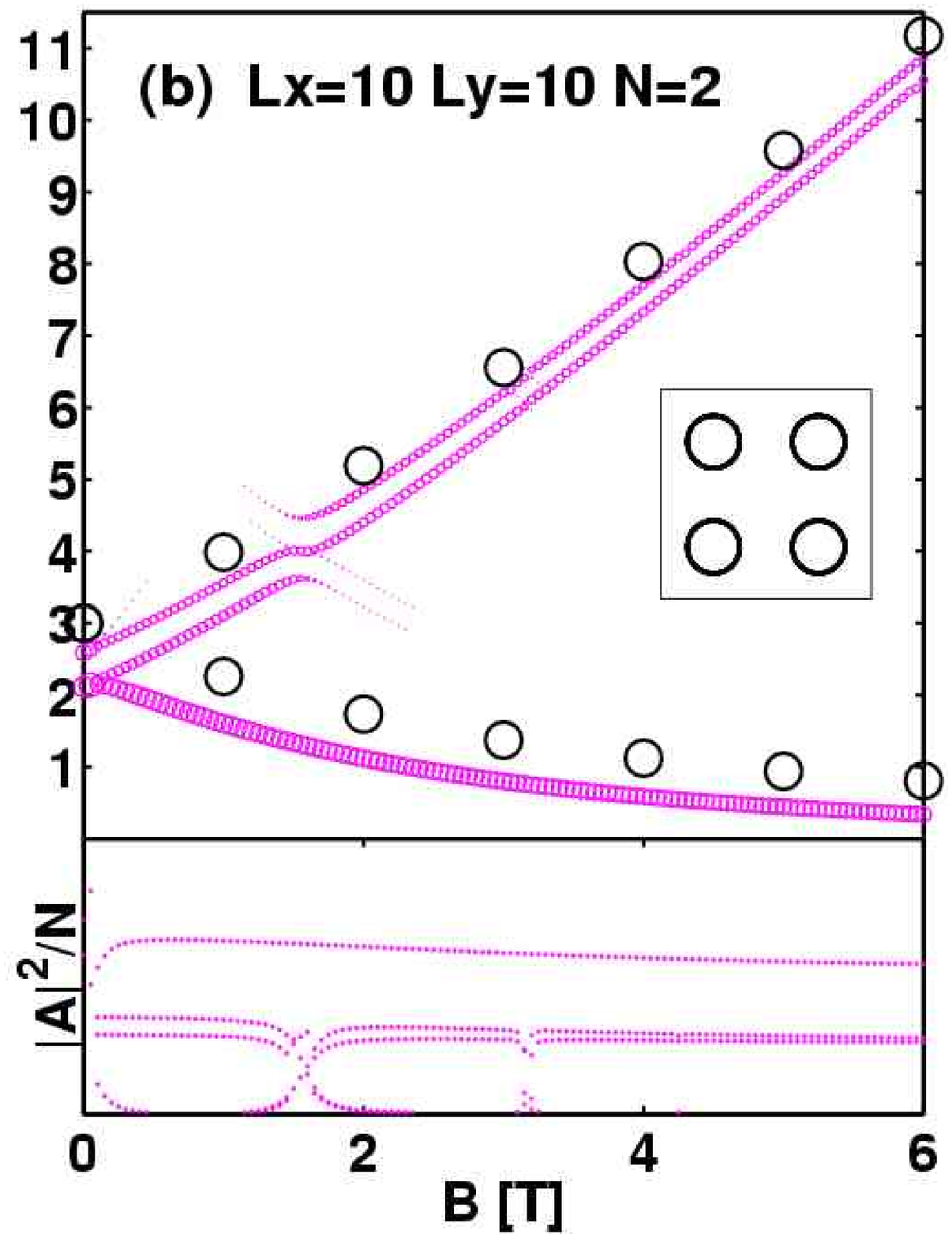}
\includegraphics*[width=0.32\columnwidth]{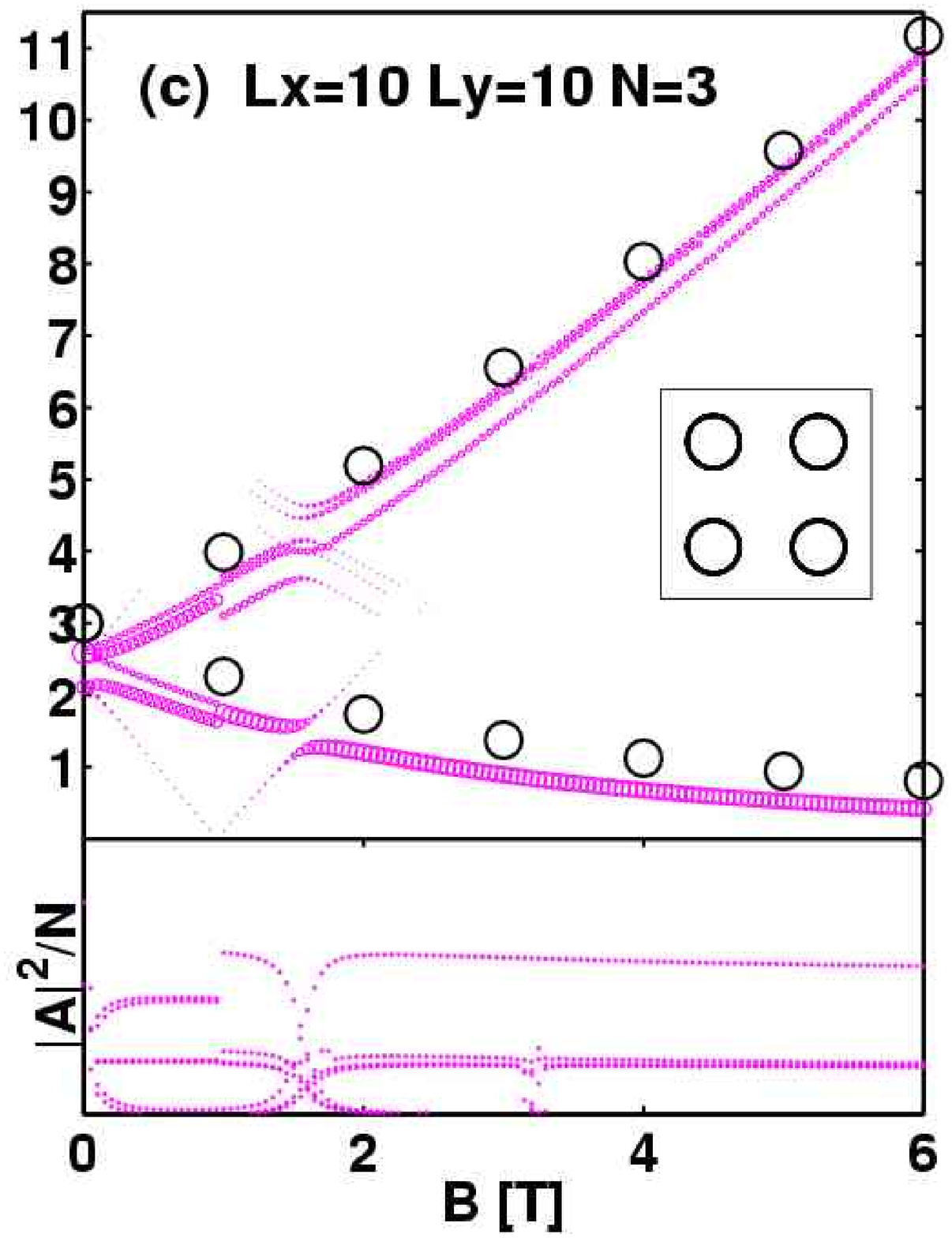}

\hfill
\includegraphics*[width=0.32\columnwidth]{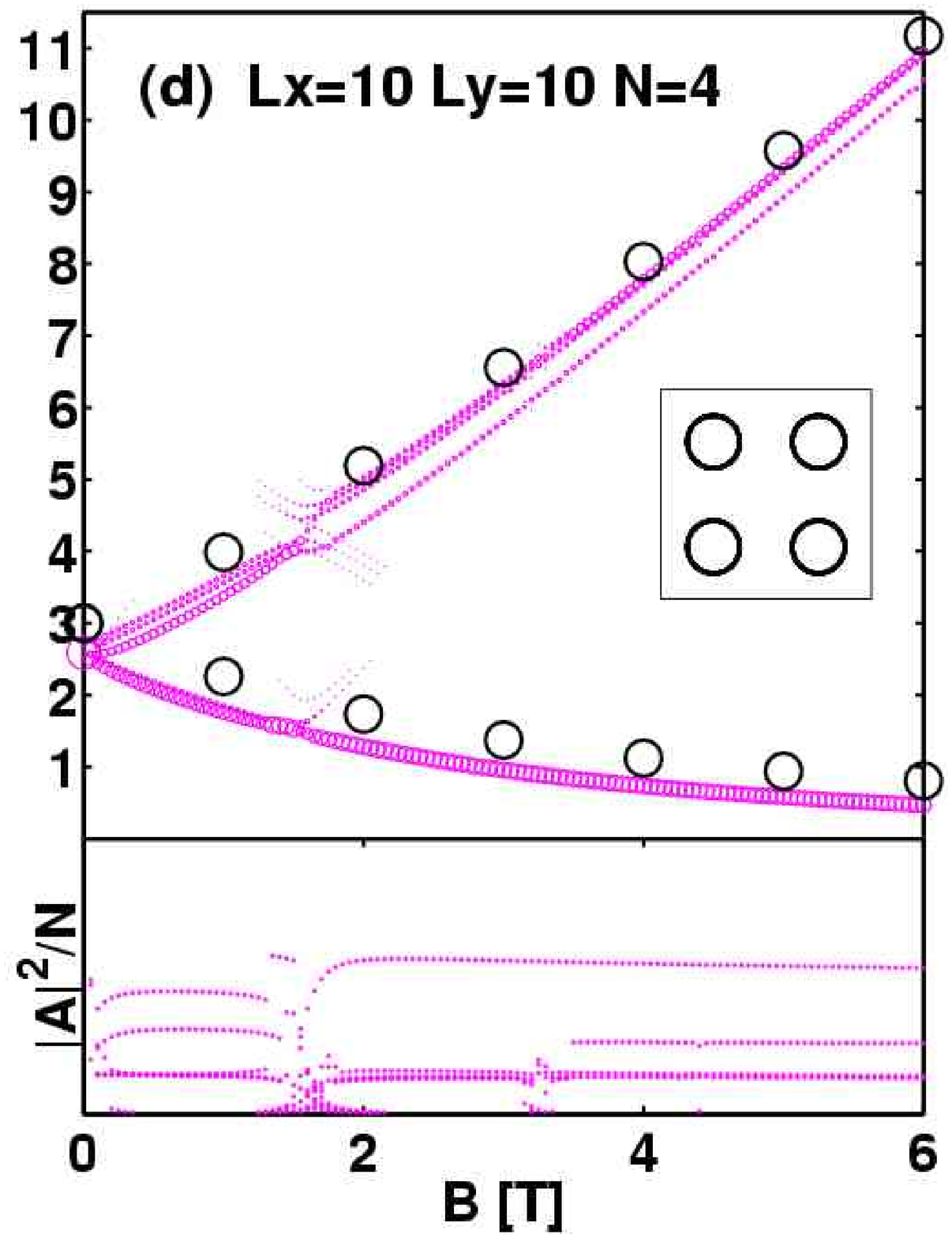}
\includegraphics*[width=0.32\columnwidth]{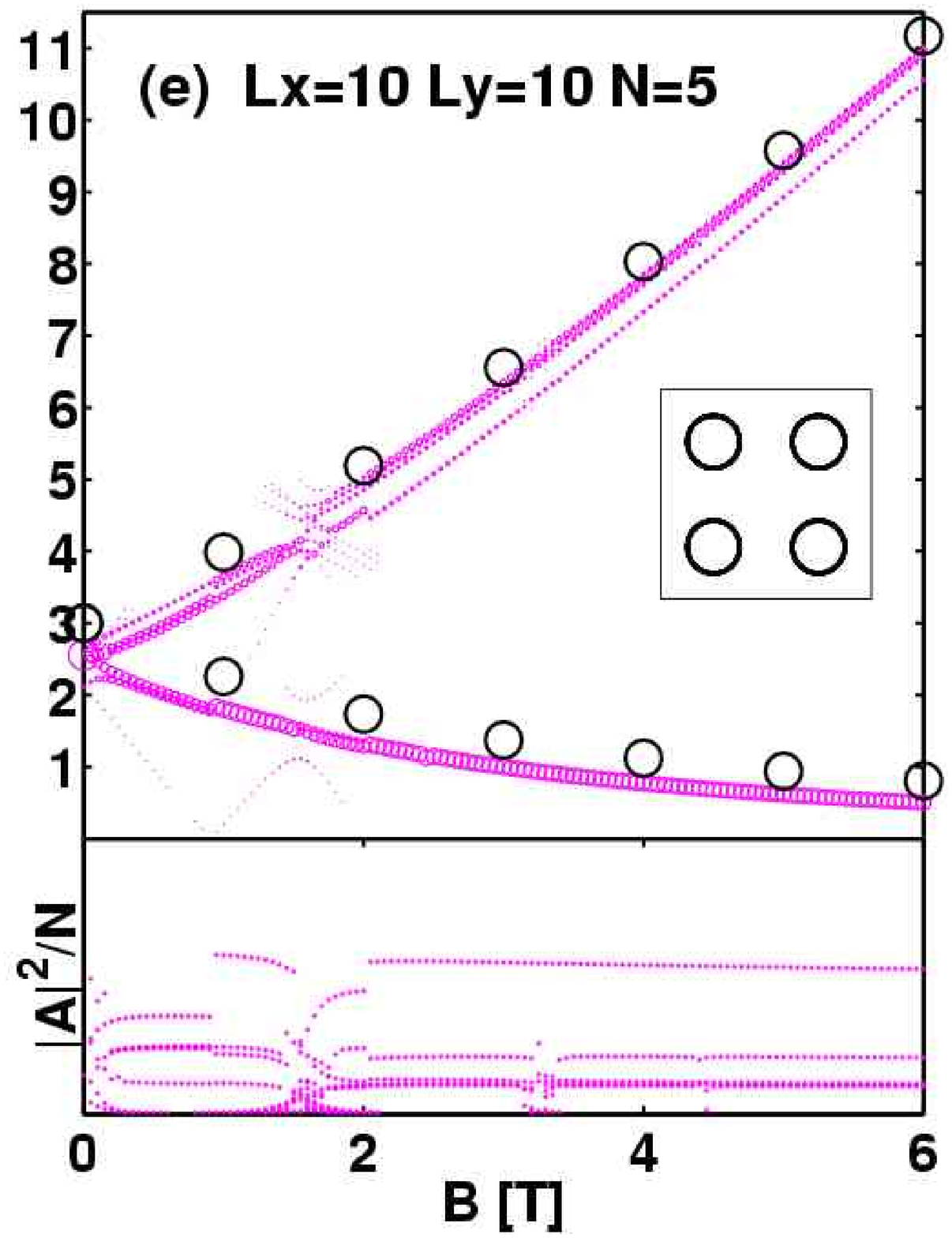}
\includegraphics*[width=0.32\columnwidth]{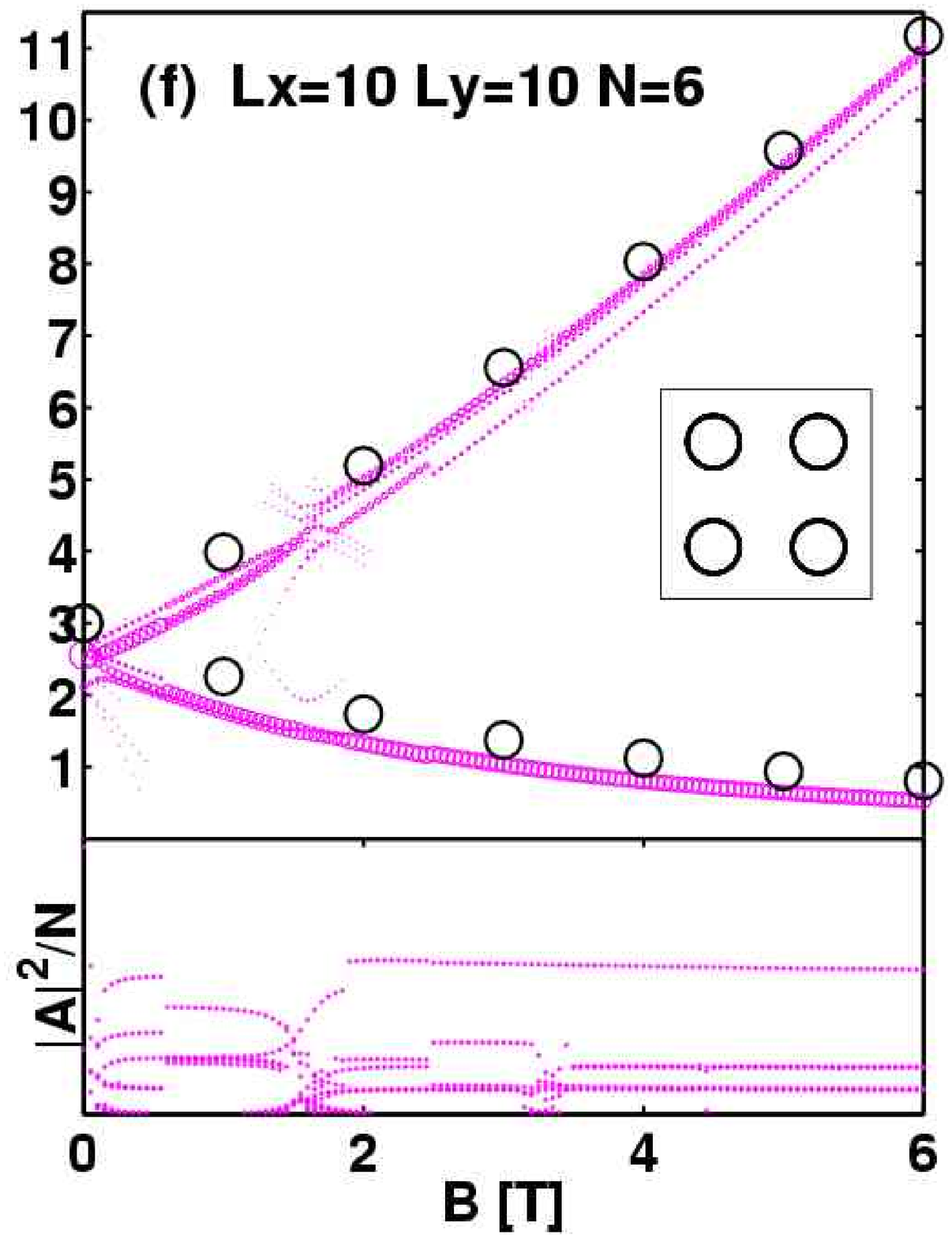}
\caption{Far-infrared spectra for $N=1-6$ non-interacting electrons in
square-symmetric four-minima quantum-dot molecule ($L_x=L_y=10$ nm).}
\label{YKSLx10Ly10}
\end{figure}

\begin{figure}
\hfill
\includegraphics*[width=0.32\columnwidth]{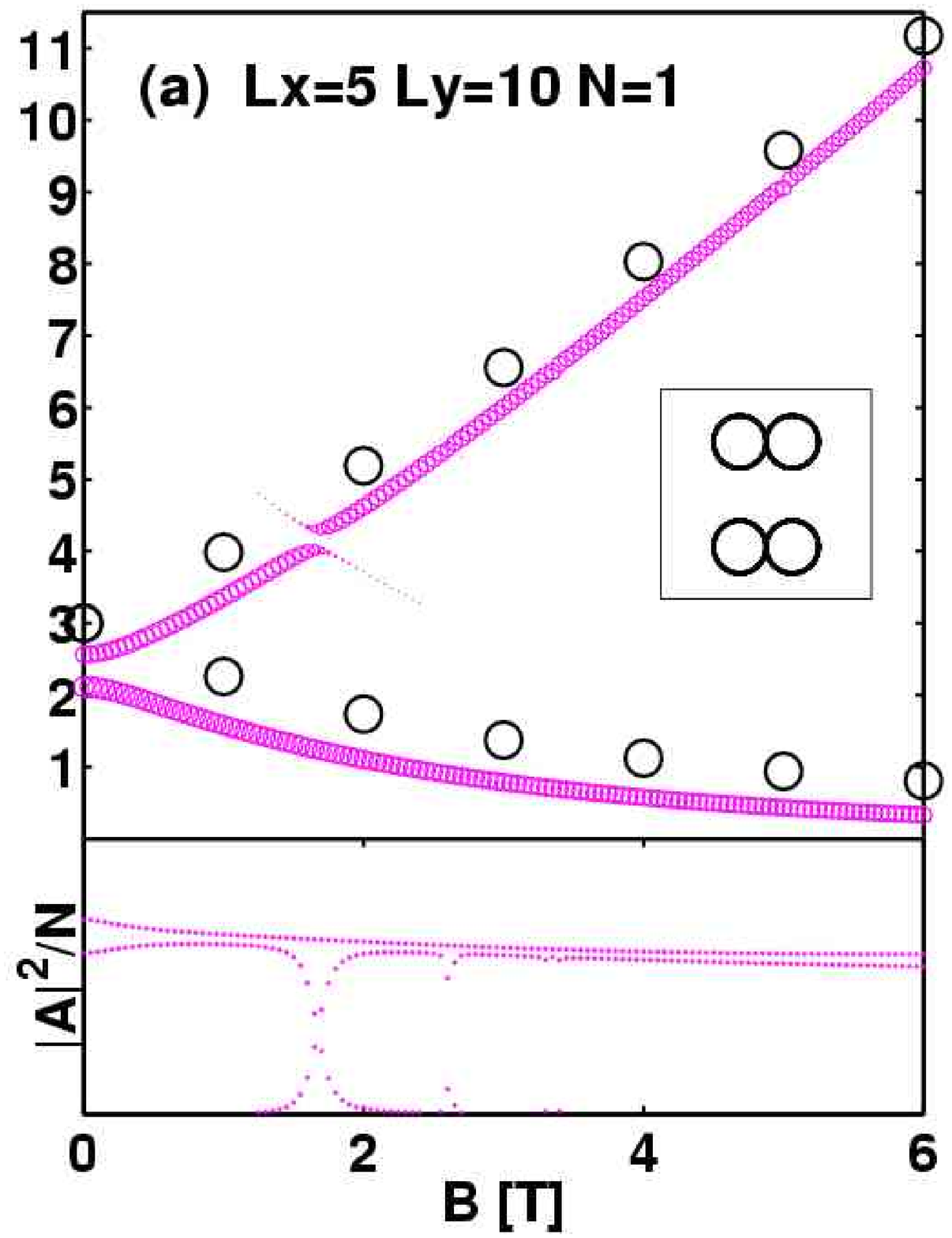}
\includegraphics*[width=0.32\columnwidth]{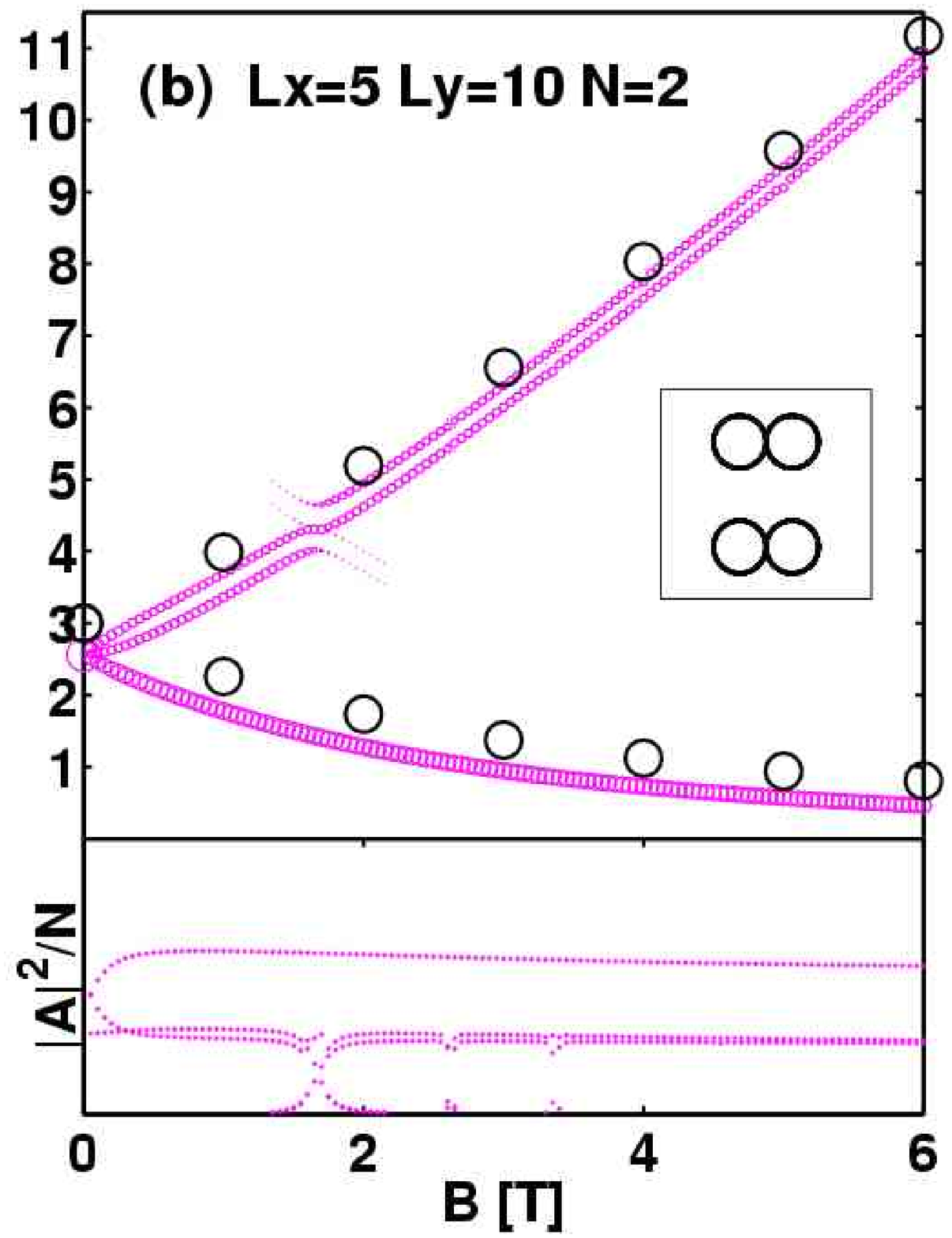}
\includegraphics*[width=0.32\columnwidth]{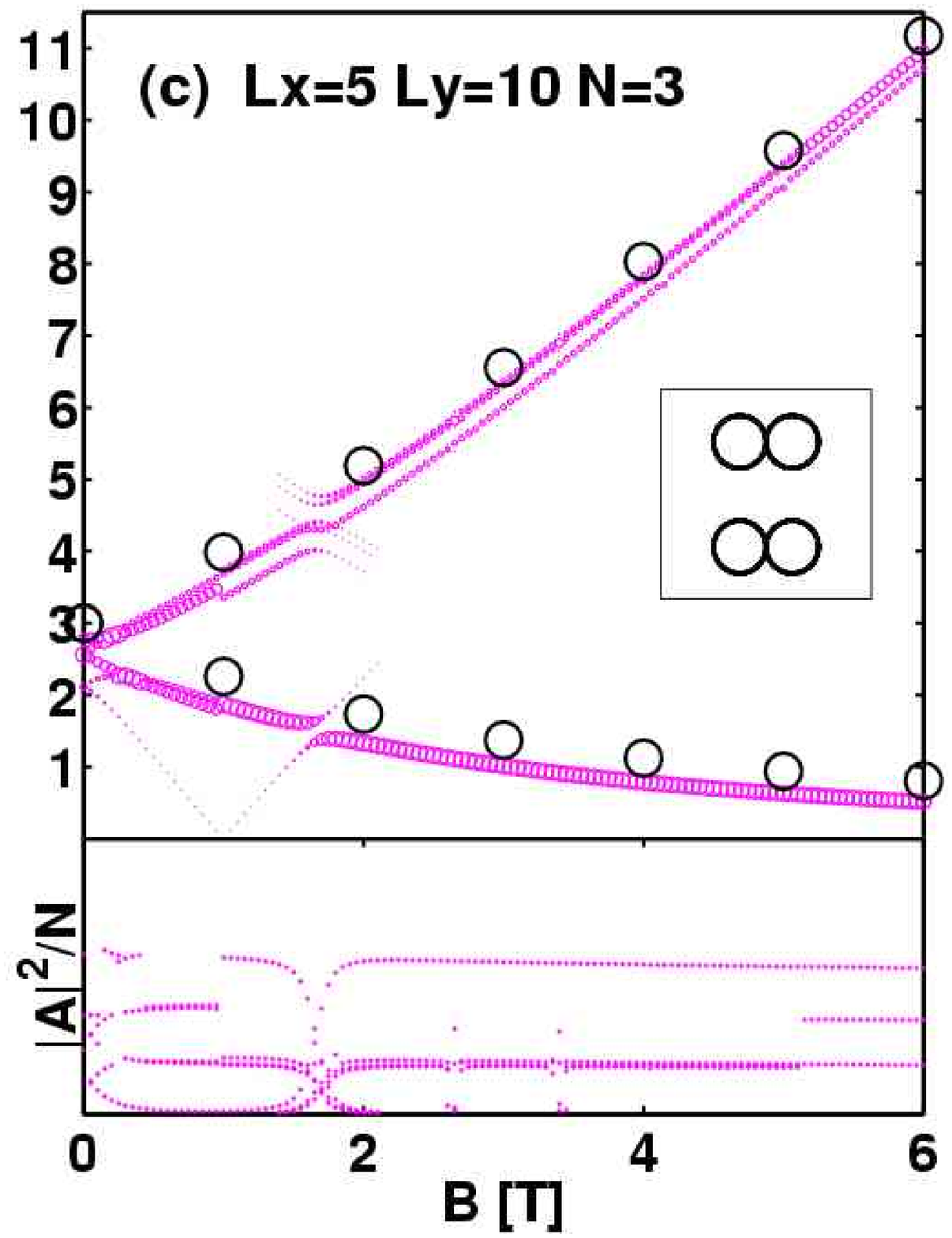}

\hfill
\includegraphics*[width=0.32\columnwidth]{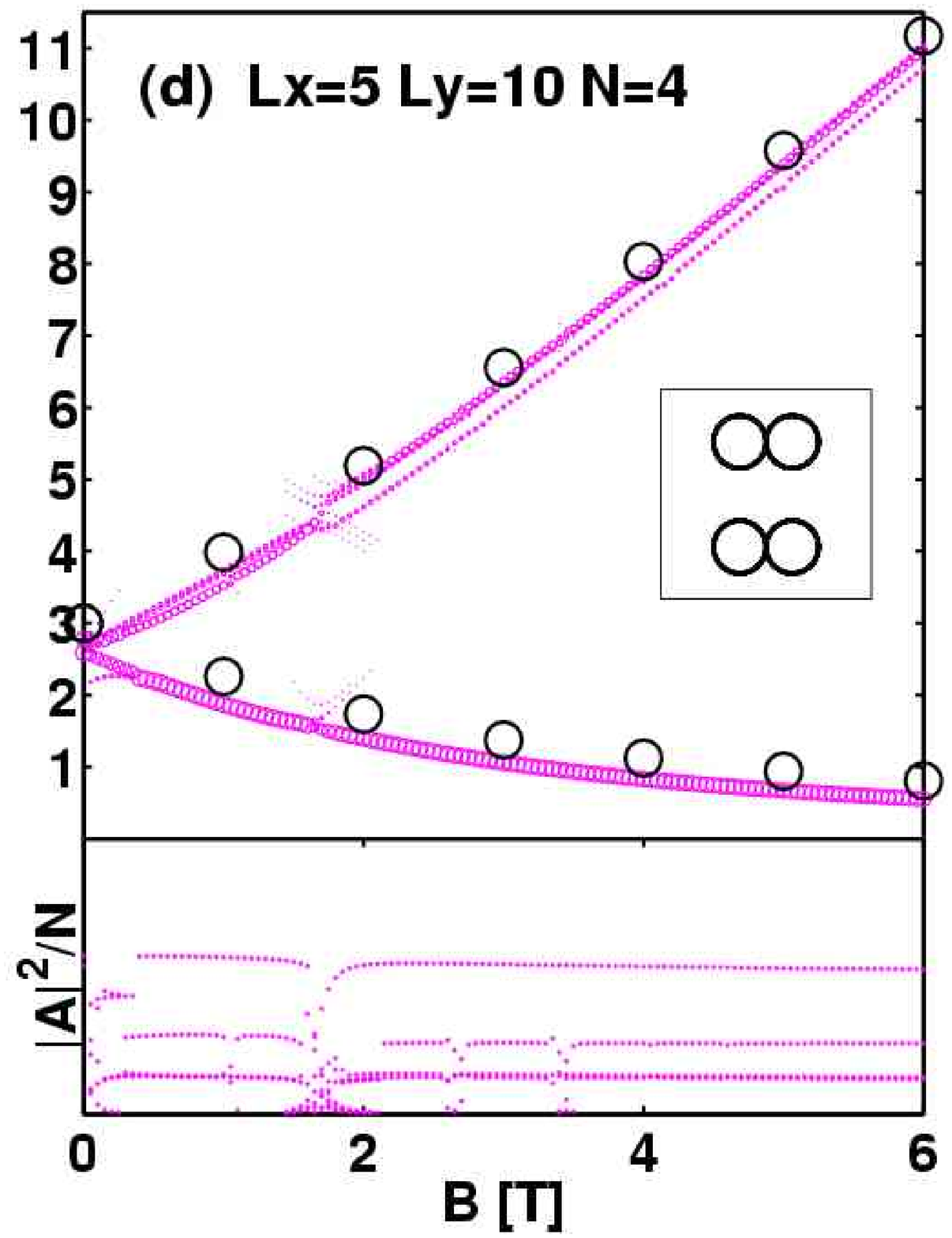}
\includegraphics*[width=0.32\columnwidth]{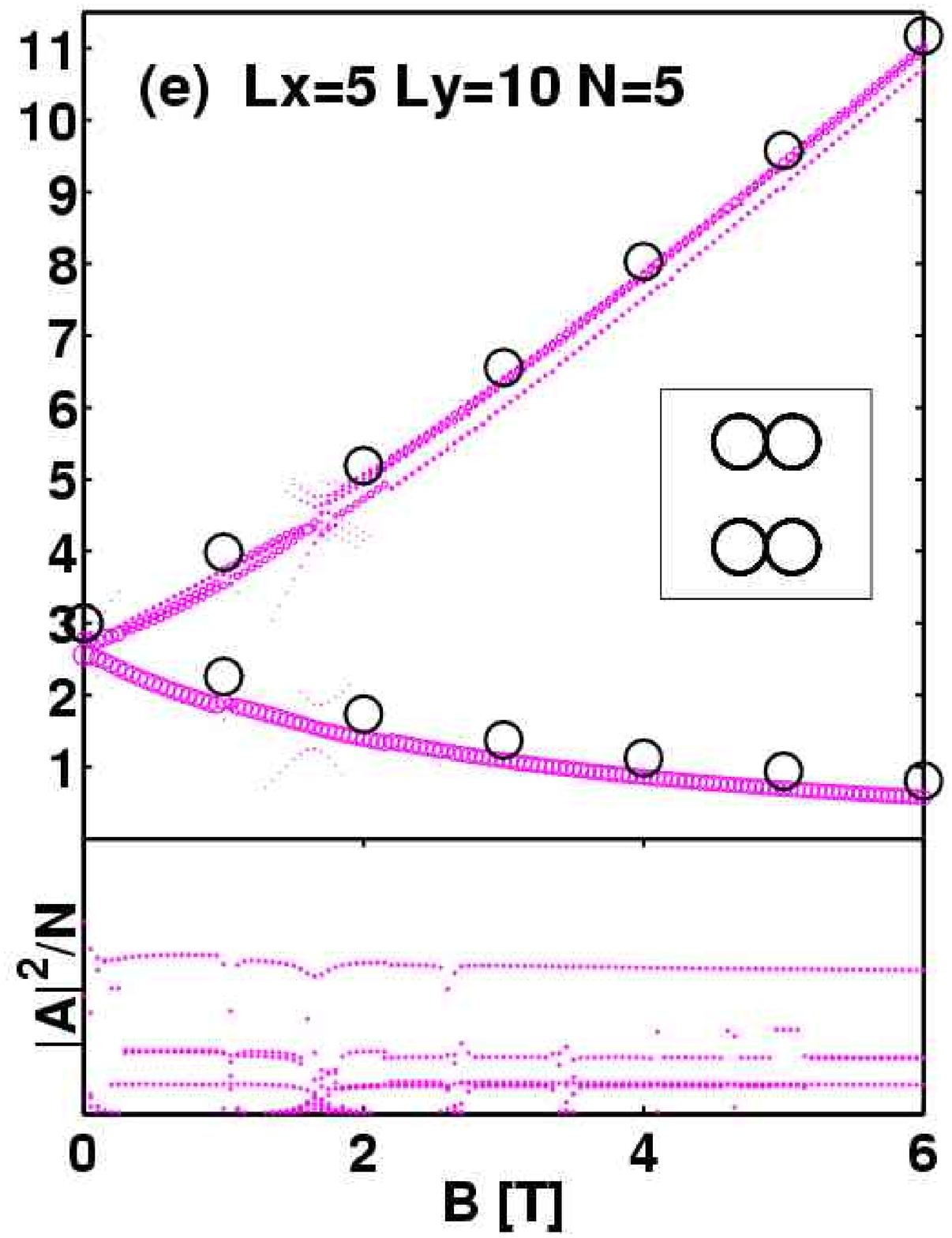}
\includegraphics*[width=0.32\columnwidth]{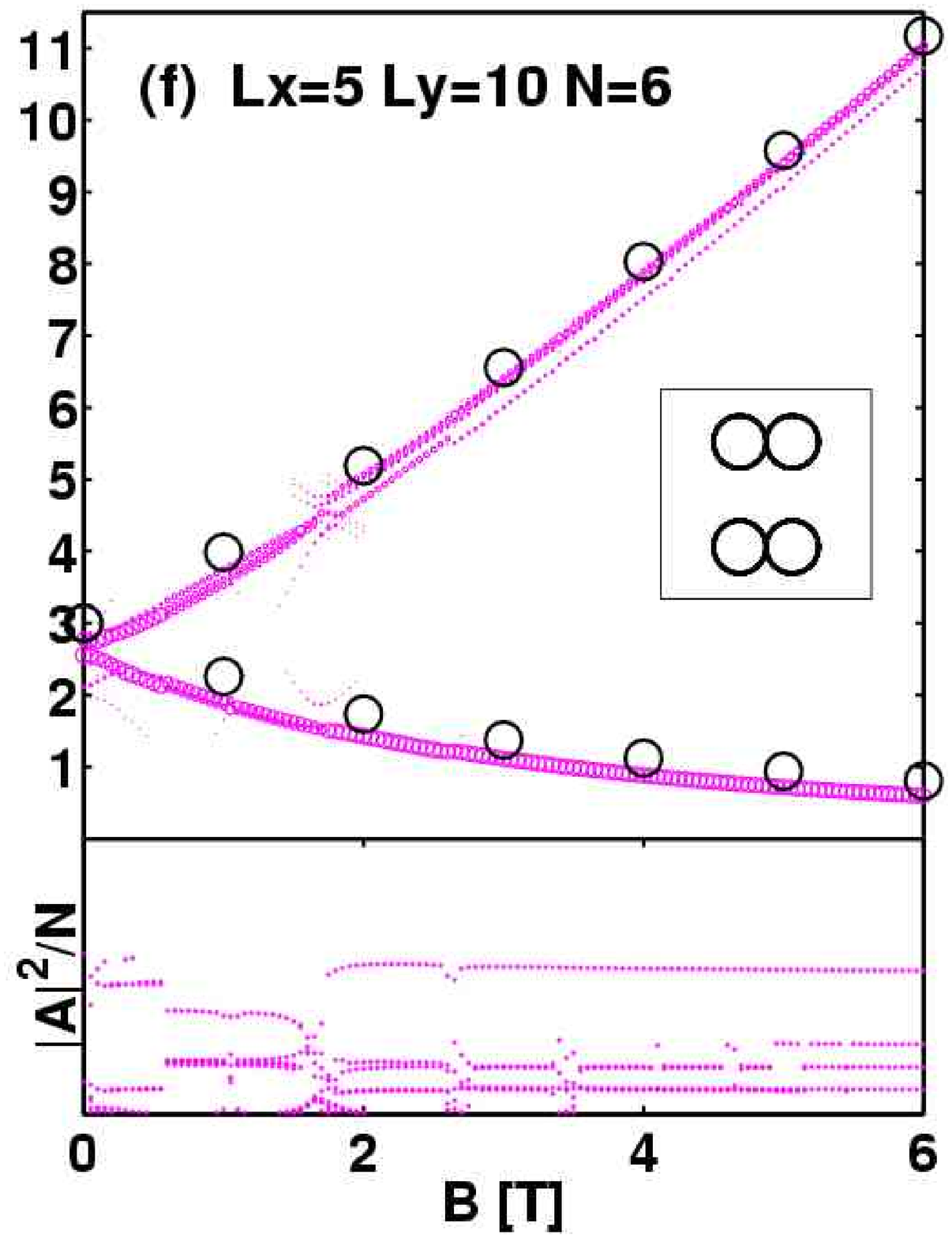}
\caption{Far-infrared spectra for $N=1-6$ non-interacting electrons in
rectangular four-minima QDM ($L_x=5,L_y=10$ nm).}
\label{YKSLx5Ly10}
\end{figure}

We obtain noninteracting FIR spectra of QDMs by occupying lowest
single-particle energy levels with $N$ non-interacting electrons and
calculating dipole transitions to higher unoccupied levels.
The noninteracting FIR spectra are shown in Figs.~\ref{YKSLx15Ly0},
~\ref{YKSLx10Ly10} and~\ref{YKSLx5Ly10} for a double dot,
square-symmetric four-minima QDM and rectangular-symmetric four-minima
QDM, respectively. (a) to (d) correspond to FIR spectra of $N=1$ to
$N=6$ electrons confined in a QDM. $N=1$ and $N=2$ are also shown
in Figs.~\ref{FIRS0} and~\ref{FIRS1}. The transition probabilities
are multiplied when more electrons are added in the QDM. We divide
each transition probability with $N$ in order to ease comparisons.

We will first analyze the single-particle spectra of a double
dot. Fig.~\ref{YKSLx15Ly0} (c) for $N=3$ shows an additional mode
below $\omega_-$ in the low-field region with the excitation energy
below $E=2$ meV. Also another mode is visible with a small transition
probability which has zero excitation energy at $B \approx 0.9$ T. If
we study the single-particle energy levels of Fig.~\ref{ele_yks} (a),
one can see that with the three lowest levels occupied, the uppermost
levels cross at $B \approx 0.9$ T. When these levels cross, the weak
mode has excitation energy of $\Delta E = 0$ and the transition
probability with the $E \leq 2$ meV mode vanishes.  Another feature
that is observed with $N=3$, but absent in two-particle spectra, is an
anticrossing in $\omega_{-}$. This anticrossing is opposite to the
anticrossings in $\omega_{+}$ where the low-field mode curves upwards
and the high-field mode curves from down to up while increasing its
strength. $\omega_+$ with three electrons looks quite similar as
$\omega_+$ in $N=2$ spectrum.

In $N=4$ spectrum of Fig.~\ref{YKSLx15Ly0} (d) there is one level
below $\omega_-$ at low $B$ but the clear anticrossing, present in
$N=3$, is missing from the spectrum. After $B \approx 1.7$ T there is
just one lower mode. The $N=5$ and $6$ spectra are rather featureless
as clear anticrossings and additional modes are missing. There are
many different levels but as they are lying close in energy, the
overall spectrum resembles just two Kohn modes without clear
additional features.  In general, anticrossings are smaller and fewer
clearly discernible additional modes are visible with $N=5$ and $6$.
Actually, there is some structure of $\omega_-$ of $N=5$ and $N=6$ but
as energies lie so close, the structure would be very difficult to
observe experimentally.  However, if the perturbation from parabolic
potential is stronger these levels may have larger energy separation
as in the experiments of Hochgr\"afe
\etal~\cite{HochgrafePRB00}. Another noticeable feature in the
noninteracting FIR spectra of double dot is the zero field gap between
the two main modes. For $N=1$ it clearly has the greatest value
whereas for other electron numbers it does not vary much.

Fig.~\ref{YKSLx10Ly10} shows non-interacting FIR spectra of
square-symmetric four-minima QDM. In $N=3$ spectrum of
Fig.~\ref{YKSLx10Ly10}(c) there is one additional mode below
$\omega_-$ and one anticrossing in $\omega_-$. The additional mode
below $\omega_-$ is much closer to $\omega_-$ in energy as it is in
the double dot spectrum. Also $\omega_-$ has lower transition
probability than the additional mode in four-minima QDM. $N=4,5$ and
$N=6$ spectra of four-minima QDM show multiple modes of $\omega_+$ and
some structure in $\omega_-$, but these are very close in energy and
may not be resolved in experiments. However, there is a discernible
split-off mode below $\omega_+$ mode with a lower transition
probability. It is difficult to say how anticrossing gaps of
$\omega_+$ at $B \approx 1.7$ T are changing with increasing $N$
because of multiple modes.  One interesting thing is that zero-field
excitation of the degenerate main branches is higher, $E \approx 2.6$ meV, with
$N=3-6$ than it is with $N=1$ and $2$, where $E \approx 2.1$ meV.

Noninteracting FIR spectra of rectangular-symmetric four-minima QDM in
Fig.~\ref{YKSLx5Ly10} does not deviate significantly from the FIR
spectra of two other QDM confinements.  Additional mode below
$\omega_-$ is very weak but visible and exist only at very low
magnetic field strengths in rectangular-symmetric four-minima QDM for
$N=3,4$ and $N=6$ in Fig.~\ref{YKSLx5Ly10}.  Clear anticrossing in
$\omega_-$ occurs only with $N=3$. The zero-field gap between
$\omega_+$ and $\omega_-$ becomes very small with with more than one
electron in the QDM. After the first anticrossing, at $B \approx 2$ T,
the $\omega_+$ is split to two modes with $N \geq 1$, where the lower
mode has lower transition probability.

\section{Summary and conclusions}
\label{Summary}

In the calculated far-infrared (FIR) spectra we are able to compare
three different quantum-dot molecule (QDM) confinement potentials and
study the effects of interactions on the two-electron FIR spectra. We
have also calculated FIR spectra for a few noninteracting
electrons. The three studied lateral QDM confinements are a two-minima
QDM (double dot), and square-symmetric and rectangular-symmetric
four-minima QDMs.

We observe anticrossings in the upper $\omega_+$ mode for both interacting and
noninteracting spectra. The anticrossing gaps are not significantly
altered by the interactions, but the anticrossing positions are
shifted to lower magnetic field values with the interacting
electrons. When we increase the particle number in the noninteracting
FIR spectra, the anticrossings in $\omega_+$ modes become less clear as
the main mode is split to multiple modes at the anticrossing point.
With three noninteracting electrons ($N=3$) we see a clear
anticrossing also in the lower $\omega_-$ mode in all three QDM
confinement potentials.

The $\omega_+$ mode is split to two for $N \geq 2$ noninteracting
electrons and also in the interacting spectra when the spin-triplet is
the ground state. With two interacting electrons this additional mode
vanishes from the spectrum at greater magnetic field values,
approximately at $B \geq 4$ for all QDM confinements. The upper
split-off mode has a lower transition probability in the interacting
case in double dot and rectangular four-minima QDM confinements. In
the noninteracting case, even if we add electrons up to six electrons
in a QDM, the $\omega_+$ consists of only two energy-split
excitations, where one of the modes, typically the lower one, can have
smaller transition probability in all three QDM confinements.

The third point that we have analyzed in the FIR spectra is the
splitting between the main modes, $\omega_\pm$, at $B=0$. The
splitting is seen for two rectangular confinements, double dot and
rectangular-symmetric four-minima QDM, where the excitations along
short and long axes are not equal. In the two rectangular
confinements, the greatest gap is seen for single electron. For all $N
\geq 2$ the gap is clearly smaller compared to $N=1$ but does not
change significantly between $N=2-6$ noninteracting electrons in the
QDM.  The zero-field splitting usually diminishes as electron-electron
interactions effectively steepen the potential which leads to increase
in excitation energies. Interactions have stronger effect on the lower
mode, which corresponds to excitation along the long axis of the
confinement, therefore decreasing the gap.

Generally, QDM confinements have lower excitation energies compared to
a parabolic quantum dot with the same confinement energy $\hbar
\omega_0$. When interactions are turned on, the Coulomb repulsion
effectively steepen the potential leading to a small increase in the
excitation energies.

The same features are in general present in interacting
and noninteracting spectra, but small shifts may occur with
interacting electrons compared to the noninteracting FIR
spectra. Therefore, based on calculations with two interacting
electrons, we conclude that deviations from the Kohn modes are due to
nonparabolic confinement potential. Interactions only shift some
features in the observed spectra but otherwise excitations are of
collective nature. However, this study is only for two interacting
electrons and maybe more features can be seen with more interacting
electrons in a nonparabolic quantum dot confinement. Also it is
possible, but unlikely, that different nonparabolic confinement
potentials could show more interaction effects in FIR spectra.

We have studied how the FIR spectra change when interactions are
turned on and off. These predictions could be checked in experiments
when the number of electrons is increased in the dots, even if the
limit of single or two electrons could not be achieved. These
predictions are that {\it i}) anticrossings are shifted to lower
magnetic field values, {\it ii}) excitation energies increase and {\it
iii}) zero field gaps decrease as the Coulomb interaction increases in
the quantum dot (more electrons inside dots). It is also possible that
the confinement potential may change as the number of electrons
changes in the dot~\cite{Henri_QDsQHE_PRL05}.  In a rectangular
symmetry also polarized light can be used to analyze
excitations. These features are seen, in general terms, in elliptic
quantum dots studied by Hochgr\"afe
\etal~\cite{HochgrafePRB00}. However, they observe a bit more
complicated FIR spectra with multiple negative and positive dispersion
Kohn modes.


This work has been supported by the Academy of Finland through its
Centers of Excellence Program (2000-2005). 
  


\providecommand{\newblock}{}

\end{document}